\documentclass[aps,pra,superscriptaddress,preprint,showpacs]{revtex4}

\usepackage{graphicx}
\usepackage{graphics}
\usepackage{amssymb}
\usepackage{amsmath}
\usepackage{epsfig}
\usepackage{latexsym}
\usepackage{color}
\usepackage{rotating}
\usepackage{subfigure}

\begin{document}

\title{Entangled Mechanical Cat States \\
via Conditional Single Photon Optomechanics}

\author{Uzma Akram}
\email{uzma@physics.uq.edu.au}
\affiliation{Centre for Engineered Quantum Systems, School of Mathematics and Physics, The University of Queensland, St Lucia, QLD 4072, Australia}
\author{Warwick P. Bowen}
\affiliation{Centre for Engineered Quantum Systems, School of Mathematics and Physics, The University of Queensland, St Lucia, QLD 4072, Australia}
\author{G. J. Milburn}
\affiliation{Centre for Engineered Quantum Systems, School of Mathematics and Physics, The University of Queensland, St Lucia, QLD 4072, Australia}

\begin{abstract}
We study single photon optomechanics conditioned on photon counting events. By selecting only detection events that occur long after a photon pulse arrives at the cavity, the optomechanical interaction time can be increased, allowing a large momentum kick to be applied to the oscillator. We apply this to two optomechanical cavities set up within a Mach-Zhender interferometer driven by a single photon source. The conditional state of the mechanical modes in such a system becomes an entangled cat state for detection times resulting in maximum mechanical amplitude in phase space. Further we  study the dynamics induced by a second photon pulse injected into an already conditioned optomechanical cavity, a quarter of a mechanical period after the first photon has been detected. We illustrate how the optomechanical interaction resulting from the second photon can be strongly suppressed allowing conditional optomechanical routing of single photons with single photon control pulses.
\end{abstract}

\maketitle
\section{Introduction}
Single photon optomechanics requires a strong coupling between optical and mechanical degrees of freedom \cite{ybchen,aspelmeyer,akram}.  If the optomechanical (OM) interaction is insufficient to reach the strong coupling regime, adding a strong coherent driving field enables the OM interaction to be enhanced due to the steady state displacement of the field amplitude in the cavity. In the strong coupling regime a single photon after entering the cavity can then be coherently exchanged between the optical and mechanical resonator. This approach falls under the "linearised" regime of optomechanics where many other novel features such as sideband cooling \cite{Wilson-Rae}, steady state optomechanical entanglement \cite{vitali,genes} and measurement of phonon number jumps \cite{gangat} among others have been unveiled; and an efficient quantum interface between optical photons and mechanical phonons \cite{Kippenberg} has been demonstrated. However in all of these proposals, operating in the linearised regime, the inherent quantum nonlinearity of the radiation pressure force acting on the nanomechanical oscillator is negligible. 

Following the achievement of ground state cooling of an engineered mechanical resonator \cite{connell,teufel,painter}, we anticipate the development of systems for which the bare single photon coupling rate \cite{painter,teufel} is large enough to make linearisation unnecessary. For example, transmissive optomechanics whereby multi element membranes are implemented as scatterers in a Fabry Perot cavity has been suggested as a possible route to achieve such desirable coupling strengths \cite{xuereb}. A number of authors have also recently theoretically investigated the behaviour of quantum OM systems with a large single OM coupling  photon radiation pressure force, predicting photon blockade \cite{rabl}, mechanical  nongaussian steady states \cite{nunnenkamp}, exploring effects of quantum noise on the quantum states of OM systems in the steady state \cite{bing} as well as a comprehensive analysis of photon statistics of OM systems in the nonlinear regime with coherent driving \cite{jieo-qiao,xun-wei2,kronwald,xun-wei}. Such progress in the field provides an exciting avenue to probe the quantum to classical transition beyond the atomic scale and promises to fulfil earlier predictions of creating macroscopic superpositions \cite{bose,marshall}. In this respect single photon OM interferometry utilising nonlinear radiation pressure effects has also been proposed to achieve quantum superpositions at the macroscopic scale, \cite{bouwmeester,ybchen2}. The scheme proposed in \cite{bouwmeester} introduced the idea of postselecting on long single photon detection times and thus probabilistically enhancing the interaction time of the single photon with the mechanical resonator. Conditioning on photon detection  has also recently appeared in the literature as a means of orthogonalising quantum states in the OM framework \cite{vanner2}. 

 In this work, we further explore single photon OM coupling in the nonlinearised regime, using the postselection idea from \cite{bouwmeester} and further probing the system with two consecutive single photon injections as first investigated recently without postselection in \cite{Liao}. The results presented in~\cite{bouwmeester}, considered only a very weak single photon interaction with the mechanical resonator, so that the excitation of the mechanics could be constrained by restricting its Hilbert space to just one phonon. 
In our work, we do not make this approximation: we consider all possible orders of the displacement induced on the mechanical mode from the OM interaction. Hence our results extend previous analysis of postselection in optomechanics to include the effect of nonlinear radiation pressure coupling resulting from single photon driving of an OM cavity in the strong coupling regime. We show how postselecting on long detection times for the driving photon can lead to effectively enhanced OM interaction times and nonclassical mechanical states. Specifically we consider conditionally driving two OM cavities arranged as in a Mach-Zhender interferometer. Hence the arriving photon has equal probability of interacting with each OM cavity. A similar arrangement has been implemented previously to achieve entanglement between vibrational modes in diamonds albeit without postselection \cite{walmsley}. However our results show that the conditional state of such a set-up approaches a perfect mechanical cat state when detection time is far longer than the source cavity decay rate and close to half of the mechanical period with the mean amplitude in phase space of the resonator maximised. This quantifies for the first time, the capacity of postselection to generate entangled mechanical cat states as first predicted in \cite{bouwmeester}. Our approach retains the nonlinear radiation pressure coupling at the single photon level and therefore includes scenarios where the OM cavities are not strongly pumped \cite{Li,borkje}. 

We further extend these results by analysing the conditional photon count rate for a second photon with a delayed injection after detection of the first photon. We show how injecting the second photon after a delay of a quarter cycle of the mechanical period allows the system to behave as a periodic single photon router. 

Consistent with previous approaches \cite{bouwmeester, ybchen2, Liao}, we neglect mechanical dissipation in our calculations. In experiments this approximation is justfiable for a sufficiently cold thermal bath satisfying $N_{bath}\ll Q$ where the bath phonon occupancy $N_{bath}=k_{B} T_{bath}/\hbar \omega_{m}$ for a given mechanical frequency $\omega_{m}$ and bath temperature $T_{bath}$. This requirement can be fulfilled by large mechanical quality factors so that the effect of phonons entering the oscillator from the bath is negligible over the mechanical period. To take a specific example, SiN strings have been shown to have Q as high as $7\times 10^{6}$ for a low resonance frequency of 176kHz \cite{schmid}. If cooled to a temperature of $300mK$, they would have a phonon occupancy of $N_{bath}=2.8\times 10^{4}$, far below their Q.

A unifying motivation in emerging literature has been a drive towards enhancing the nonlinearity in OM systems. Notable examples include the use of multiple optical modes in membrane in the middle set-ups to increase both their linear \cite{painter2} and quadratic \cite{jharris1,jharris2} nonlinearity, and the engineering of phononic-photonic crystals with large degrees of overlap between photonics and phononic modes \cite{painter}. By contrast, here, rather than developing a new OM configuration, a stronger OM interaction is achieved by conditionally selecting trajectories where the photon interacts for an extended period with the mechanical oscillator.

\section{Single photon conditional optomechanics}
\label{theorymodel}

\subsection{The Model}
\begin{figure}[!ht]
\centering
\includegraphics[scale=0.6]{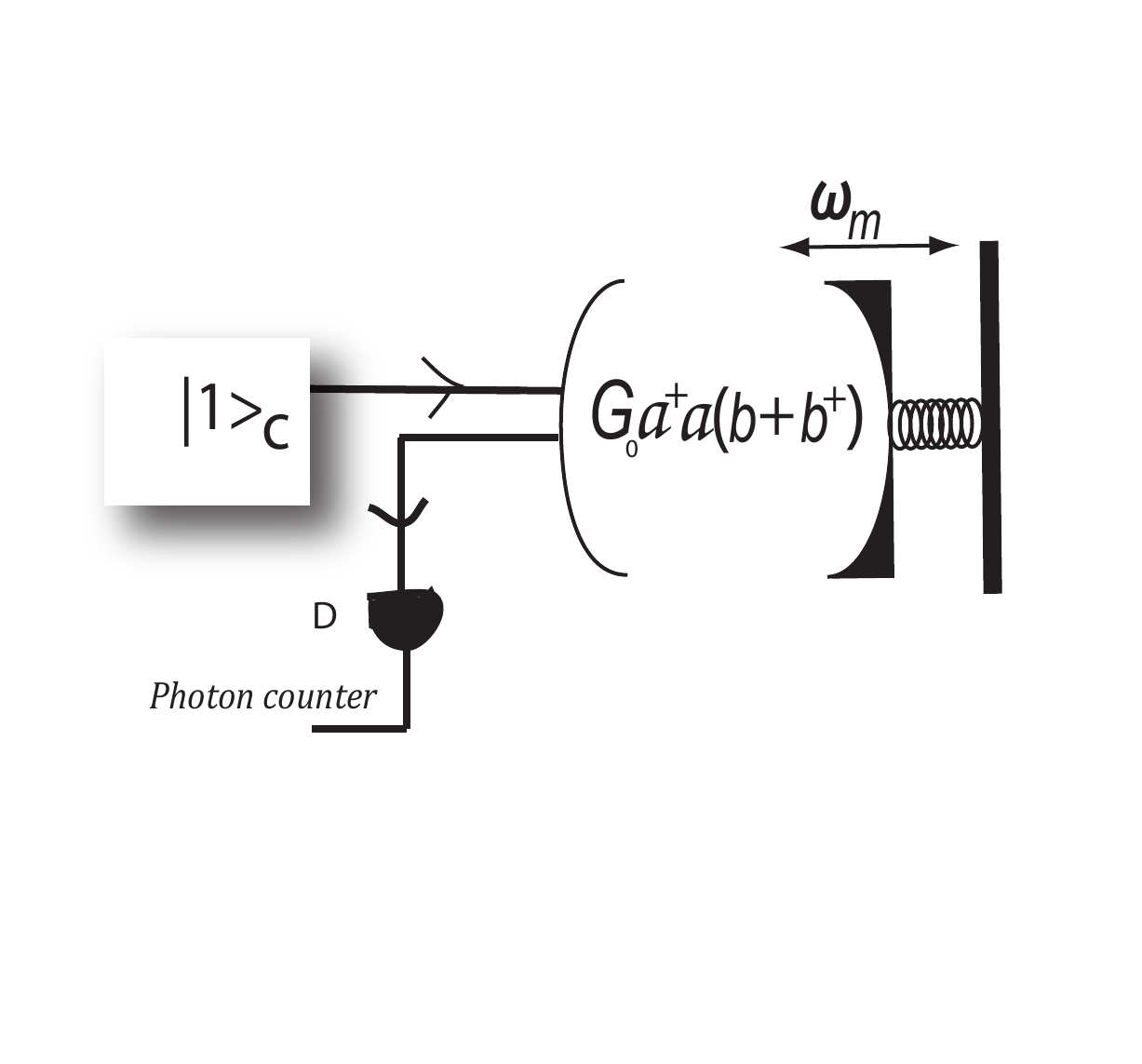}
\caption{A single sided OM cavity driven by a single photon source. Photon emissions from the composite system are monitored by a perfect photon counter D.} 
\label{model}   
\end{figure}

We consider a single sided OM cavity driven by a single photon source. The source is modelled as an independent cavity with decay rate $\gamma$ prepared at t=0 in a single photon state, and is coupled to the OM cavity irreversibly via the cascaded systems approach \cite{Car,Gar}. Photon emissions from the composite system are monitored with a single photon counter, $D$. Our set-up is summarised in Fig.\ref{model}.  

The Hamiltonian for the OM cavity, in an interaction picture at the cavity frequency, is 
\begin{equation}
H_{om}=\hbar\omega_m b^\dagger b+\hbar G_0 a^\dagger a (b+b^\dagger)
\label{h_om}
\end{equation}
where $a,a^\dagger$ are the annihilation and creation operators for the optical resonator and $b,b^\dagger$ are the annihilation and creation operators for the mechanical resonator with frequency $\omega_m$. The master equation for the cascaded source cavity and the OM system is 
\begin{equation}
\frac{d\rho}{dt} =-\frac{i}{\hbar}[H,\rho]+{\cal D}[J]\rho
\label{ME}
\end{equation}
where 
\begin{equation}
H= H_{om}+H_{cas}
\label{Htot}
\end{equation}
with 
\begin{equation}
H_{cas} =-i\sqrt{\kappa\gamma}(ca^\dagger-c^\dagger a)/2
\end{equation}
and the jump operator is given by 
\begin{equation}
J=\sqrt{\gamma}c+\sqrt{\kappa}a
\end{equation}
and $c,c^\dagger$  are the annihilation and creation operators for the source cavity, with decay constant $\gamma$. The decay rate of the optical resonator for the OM system is $\kappa$.

There are two indistinguishable temporal histories corresponding to the photon counting event at $D$: the photon can be reflected off the OM cavity directly into the photon counter without interacting with the mechanical resonator at all, or the photon can be detected after emission from the cavity having first been transmitted at the entrance mirror. We will label these two temporal histories $R$ and $T$ respectively.  From the viewpoint of the quantum trajectory theory for cascaded systems, a photon count at $D$ allows the description of the state of the system to be updated through application of the jump operator $J=\sqrt{\gamma} c+\sqrt{\kappa}a$.  This means that the resulting (unnormalised) conditional state is a superposition of the form $|\tilde{\Psi}^{(1)}\rangle = \sqrt{\gamma}|R\rangle+\sqrt{\kappa}|T\rangle$ (we do not label the time of detection at this point).  Note that throughout this paper tildes are used to signify that a state is unnormalised. As these two histories are indistinguishable we would expect to see an interference term in the single photon count rate. 

In the case of two-photon driving considered in section \ref{2photon} there will be four indistinguishable histories, $RR,RT,TR,TT$ leading to two photon counting events and two applications of the jump operator.  The resulting conditional state will hence be a superposition of these four possibilities,  $|\tilde{\Psi}^{(2)}\rangle =\gamma |RR\rangle+\kappa|TT\rangle+ \sqrt{\gamma\kappa}(|RT\rangle+|TR\rangle)$. 

\subsection{One photon conditional optomechanics}

We first consider the case where one photon drives the OM system and calculate the conditional state of the mechanical resonator given that no photon is counted up to time $t$ and then exactly one photon is counted between $t$ and $t+dt$. As only a single photon exists in the system, a click at the photon counter conditionally provides $100\%$ efficiency in the protocol even in the presence of large losses. Moreover, as there is no other channel for the photon to be lost in our model, if the initial state of the system is pure, the conditional state of the system at any time will also be pure. In particular the unnormalised conditional state of the system, $|\tilde{\Psi}^{(0)}\rangle$ given no count up to time $t$ \cite{car_notes} is given by 
\begin{equation}
\label{psi}
|\tilde{\Psi}^{(0)}(t)\rangle =\exp\left [-\frac{i}{\hbar}Ht-\frac{1}{2} J^\dagger J t\right ]|\tilde{\Psi}(0)\rangle
\end{equation}
Which implies 
\begin{equation}
\label{non-hermitian}
\frac{d |\tilde{\Psi}^{(0)}(t)\rangle}{dt} = -i\left (H/\hbar-\frac{i}{2} J^\dagger J\right )|\tilde{\Psi}^{(0)}(t)\rangle  
\end{equation} 

The initial state of the system is
\begin{equation}
\label{ansatz}
|\tilde{\Psi}(0)\rangle =|1\rangle_c|0\rangle_a|\psi\rangle_b
\end{equation}
where $|\psi\rangle_b$ is an arbitrary coherent state of the mechanical oscillator with amplitude $\psi$ which we take to be real. As there is at most one photon in the entire system at any time we can expand the unnormalised conditional state of the system as 
\begin{equation}
|\tilde{\Psi}^{(0)}(t)\rangle=|1\rangle|\phi_1(t)\rangle_{b}+|2\rangle|\phi_2(t)\rangle_{b}
\end{equation}
where we have defined 
\begin{eqnarray}
|1\rangle & = & |0\rangle_c|1\rangle_a\\
|2\rangle & = & |1\rangle_c|0\rangle_a
\label{possibilities}
\end{eqnarray}
as the photon may either be in the source or the OM cavity at any given time, $t>0$ prior to being counted.

To solve for the evolution of the system, we first note that the OM Hamiltonian in Eq.(\ref{h_om}) can be diagonalised by making the polaron transformation 
 \begin{equation}
 \bar{H}= SH S^\dagger=\omega_m b^\dagger b -\frac{G_0^2}{\omega_m} (a^\dagger a)^2
 \end{equation}
 where 
 \begin{equation}
 S= e^{\beta a^\dagger a(b-b^\dagger)}.
 \end{equation}
  where $\beta=-G_0/\omega_m$.
 This can be used to obtain a solution to the no-jump conditional evolution of Eq.(\ref{non-hermitian}). First we make the transformation to what we will call the {\em displacement picture}
 \begin{equation}
 |\tilde{\Psi}^{(0)}(t)\rangle_D = S |\tilde{\Psi}^{(0)}(t)\rangle
 \end{equation}
 for which we find 
 \begin{equation}
 \frac{ |\tilde{\Psi}^{(0)}(t)\rangle_D}{dt} =[-i\bar{H}-\frac{1}{2} \bar{J}^\dagger \bar{J}] |\tilde{\Psi}^{(0)}(t)\rangle_D
 \end{equation}
 with 
 \begin{eqnarray}
 \bar{H} &=&\omega_m b^\dagger b -\frac{G_0^2}{\omega_m} (a^\dagger a)^2-i\sqrt{\kappa\gamma}(ca^\dagger D(\beta)-c^\dagger a D^\dagger(\beta))/2 \\
 \bar{J}&=&\sqrt{\gamma}c +\sqrt{\kappa}a e^{\beta (b^{\dagger}-b)}
\end{eqnarray}
where $D(\beta)$ is a displacement operator. 

We now transform the ansatz in Eq.(\ref{ansatz}). In the displacement picture it becomes, 
\begin{eqnarray}
  |\tilde{\Psi}^{(0)}(t)\rangle_D & = &  |1\rangle|\bar{\phi}_1(t)\rangle_{b}+|2\rangle|\bar{\phi}_2(t)\rangle_{b}\\
  & = &  |1\rangle D(\beta) |\phi_1(t)\rangle_{b}+|2\rangle|\phi_2(t)\rangle_{b}
  \end{eqnarray}
  thus the only modification in the displacement picture is that
  \begin{equation}
  |\bar{\phi}_1(t)\rangle_{b}= D(\beta) |\phi_1(t)\rangle_{b}
  \end{equation}

  At $t=0$ the photon is definitely in the source cavity, so that the initial state of the system can be specified to be 
  \begin{equation}
    |\tilde{\Psi}^{(0)}(t_i)\rangle_D =S|2\rangle|\psi_i\rangle=|2\rangle |\psi_i\rangle
    \end{equation}
  at an initial time $t_i$, where $|\psi\rangle$ is an {\em arbitrary} mechanical state. 
  The non-detection evolution in Eq.(\ref{non-hermitian}) then becomes,
  \begin{eqnarray}
  \frac{d|\bar{\phi}_1\rangle_{b}}{dt} & = & -[i\omega_m b^\dagger b-i\chi+\kappa/2]|\bar{\phi}_1\rangle_{b}-\sqrt{\kappa\gamma}D(\beta)|\bar{\phi}_2\rangle_{b}\\
    \frac{d|\bar{\phi}_2\rangle_{b}}{dt} & = &  -[i\omega_m b^\dagger b+\gamma/2]|\bar{\phi}_2\rangle_{b}
   \end{eqnarray}
where $\chi=G^{2}_{0}/\omega_{m}$. Hence the OM interaction with the driving photon induces a Kerr-like nonlinearity in the state dynamics, which can be enhanced by a large OM coupling strength, $G_{0}$ and small mechanical frequency $\omega_{m}$.  
   We can solve the second of the equations above immediately and substitute into the first. Transforming back to the original picture from the displacement picture we find that 
    \begin{eqnarray}
    |\phi_2(t_f)\rangle_{b}  & = & e^{-(i\omega_m b^\dagger b+\gamma/2)(t_f-t_i)}|\psi_i\rangle_{b}\\
       |\phi_1(t_f)\rangle_{b}  & = & \hat{K}(t_f;t_i)|\psi_i\rangle_{b}
       \end{eqnarray}
       where the propagator is found to be     
       \begin{eqnarray}
       \label{full-prop}
  \hat{K}(t_f;t_i)  & =  &  -\sqrt{\kappa\gamma}D^\dagger(\beta) e^{-(i\omega_mb^\dagger b -i\chi+\kappa/2)(t_f-t_i)}\\
 & & \, \, \, \,  \times \int_{t_i}^{t_f} dt' e^{(i\omega_mb^\dagger b-i\chi+\kappa/2)t'}D(\beta) e^{-(i\omega_mb^\dagger b+\gamma/2)t'}
  \end{eqnarray}

Henceforth we consider the special case that the mechanics starts in the ground state, $|\psi\rangle_{b} =|0\rangle$ at $t_i=0$. In this special case we find that
    \begin{eqnarray}
     |\phi_1(t)\rangle_{b}    & = & D^\dagger(\beta)\hat{R}(t) D(\beta)|0\rangle_{b} \label{conditional-mech1}\\
|\phi_2(t)\rangle_{b} & = & e^{-\gamma t/2}|0\rangle_{b}
\label{conditional-mech2}
 \end{eqnarray}
 where 
 \begin{equation}
 \hat{R}(t)= \sum_{n=0}^\infty \frac{(e^{-\gamma t/2}-e^{-(i\omega_m n-i\chi +\kappa/2)t})}{i\omega_m n-i\chi+(\kappa-\gamma)/2}|n\rangle\langle n|
 \label{hatR}
  \end{equation}
 The unnormalised conditional state, given that no photons are counted up to time $t$ is then
\begin{equation}
|\tilde{\Psi}^{(0)}(t)\rangle = |1\rangle|\phi_1(t)\rangle_{b}+|2\rangle|0\rangle_{b}e^{-\gamma t/2}
  \end{equation}
  At any time $t>t_{i}$, the system evolves as a superposition of the two possible states, $|1\rangle$ and $|2\rangle$. This is the key idea behind generation of superposition states in this work. As $t$ increases, the probability of the photon to be in the source, $|2 \rangle$, reduces to zero. 
If the photon is counted between $t$ and $t+dt$, the resulting conditional state is found by applying the the jump operator $J$ to get the unnormalised conditional state
  \begin{equation}
  |\tilde{\Psi}^{(1)}(t)\rangle=[\sqrt{\kappa}|\phi_1(t)\rangle_{b}+\sqrt{\gamma}|0\rangle_b e^{-\gamma t/2}]|0\rangle_c|0\rangle_a
  \end{equation}
  Thus the unnormalised conditional state of the mechanics given that no photons are counted up to time $t$ and one photon is counted between $t$ and $t+dt$ is 
  \begin{equation}
  \label{conditional-count}
|\tilde{\Phi}^{(1)}(t)\rangle=  \sqrt{\kappa}|\phi_1(t)\rangle+\sqrt{\gamma}e^{-\gamma t/2} |0\rangle 
\end{equation}
where we drop the suffix $b$, such that $|\tilde{\Phi}^{(1)}(t)\rangle$ denotes the conditional mechanical state at time $t$. The first term in Eq.(\ref{conditional-count}) is given in Eq.(\ref{conditional-mech1}). In the quantum trajectory formalism \cite{Car} the normalisation of the conditional state is in fact the rate for photon counts and is determined by
\begin{equation}
\label{detection_rate}
R_1(t)=\langle \tilde{\Phi}^{(1)}(t) |\tilde{\Phi}^{(1)}(t)\rangle = \kappa\langle \phi_1(t) |\phi_1(t)\rangle +\gamma e^{-\gamma t} +\sqrt{\kappa\gamma}e^{-\gamma t/2}(\langle 0|\phi_1(t)\rangle+c.c)
\end{equation}
The first term is the rate to count photons {\em given} that they come from the OM cavity. The second term is the rate to count photons given that they come straight from the source and are reflected from the OM cavity. The last term arises due to interference between photons reflected and those transmitted from inside the OM cavity. With this interpretation we see that the mean number of photons inside the OM cavity, prior to the detection, is just
\begin{equation}
\langle a^\dagger a\rangle(t)=\langle \phi_1(t) |\phi_1(t)\rangle
\end{equation}

If we assume that the mechanics starts in the ground state, we can use Eq.(\ref{conditional-mech1}) to show that
\begin{equation}
R_1(t) = \kappa\langle \beta|\hat{R}^\dagger(t) \hat{R}(t)|\beta\rangle+\gamma e^{-\gamma t}+\sqrt{\kappa\gamma}e^{-\gamma t/2}(\langle \beta |\hat{R}(t)|\beta\rangle +c.c)
\end{equation}
where $|\beta\rangle$ is a coherent state of the mechanical mode.
The mean photon number in the OM cavity prior to detection is then
\begin{equation}
\langle a^\dagger a\rangle=\langle \beta|\hat{R}^\dagger(t) \hat{R}(t)|\beta\rangle=\sum_{n=0}^\infty e^{-|\beta|^2}\frac{|\beta|^{2n}}{n!} |r_n(t)|^2
\end{equation}
where 
\begin{equation}
\label{diagonal_R}
r_n(t)=\frac{(e^{-\gamma t/2}-e^{-(i\omega_m n-i\chi +\kappa/2)t} )}{i\omega_m n-i\chi+(\kappa-\gamma)/2}
\end{equation}
The interference term in the single photon count rate (the last term in Eq.(\ref{detection_rate})) is determined by
\begin{equation} 
\langle \beta |\hat{R}(t)|\beta\rangle=\sum_{n=0}^\infty e^{-|\beta|^2}\frac{|\beta|^{2n}}{n!} r_n(t)
\end{equation}

\begin{figure}[h!]
\centering
\includegraphics[scale=0.7]{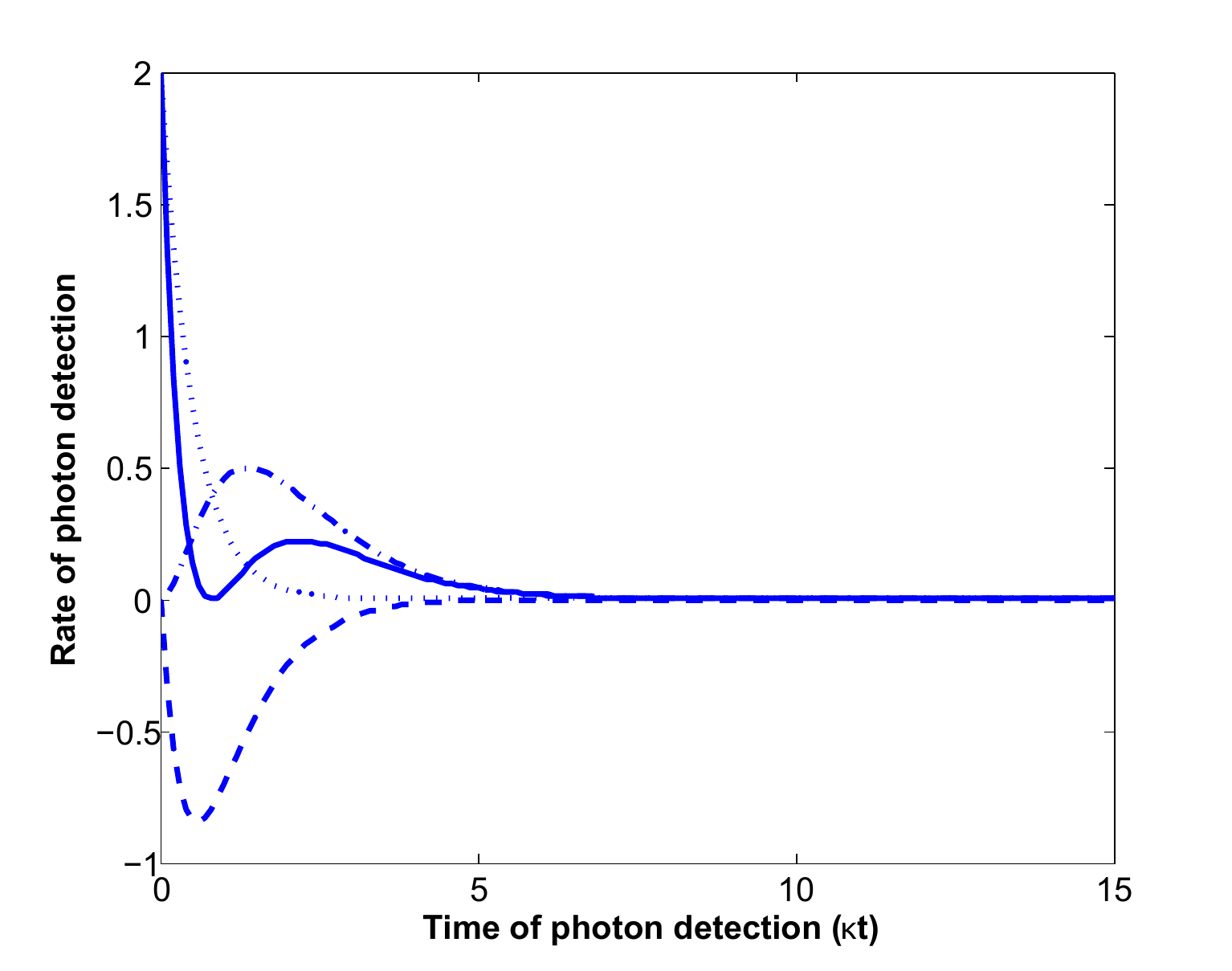}
\caption{The temporal profile of the total count rate (solid curve) of arriving photons at the photon counter $D$, which is composed of the rate of reflected photons (dotted curve), rate of  transmitted photons (dashed-dotted curve) and an interference (dashed curve) occurring between reflected and transmitted photons. Parameters are taken to be dimensionless in units of the optical decay rate $\kappa$: $\gamma/\kappa=2$, $G_{0}/\kappa=0.02$ and $\omega_{m}/\kappa=0.02$.  } 
\label{rateofphoton1}   
\end{figure}

In Fig.\ref{rateofphoton1} we show how the total rate of photon counts varies with time after the single photon source is switched on. All parameters are taken to be in units of the optical decay rate which is fixed at $\kappa$. As we are interested in enhancing the OM cooperativity, we consider a weak OM interaction of strength $G_{0}/\kappa=0.02$. In order to displace the mechanics by a large amplitude, $\beta=-G_{0}/\omega_{m}$, we choose a mechanical resonator with a small frequency, $\omega_{m}/\kappa=0.02$. As time evolves, $t>0$, the total photon count rate decays exponentially, exhibiting a minimum at a finite time. This minimum in the rate comes from the interference term in Eq.(\ref{detection_rate}), which is shown by the dashed curve in Fig.\ref{rateofphoton1}. The evolution of the mean number of photons in the OM cavity is shown as the dashed-dotted curve and corresponds exclusively to those photons which interact with the mechanical resonator.

Early detection at the photon counter will most likely correspond to photons reflected off the OM cavity, hence coming directly from the source, without interacting with the mechanical system. However detection times after the minimum in the rate will most likely correspond to photons that have interacted with the OM cavity. Hence postselection on rare late detection events ensures both that the photon entered the OM system and that it interacted for a prolonged period. In these infrequent circumstances even if the bare OM coupling is small, postselection would lead to effectively enhanced OM interaction resulting in a significant momentum kick to the mirror. These expectations can be justified by computing the moments of the conditional mechanical state at the time the photon is detected. 

\subsection{Conditional mechanical moments}
In this section we investigate the conditional state of the mechanical oscillator given a photon count between $t$ and $t+dt$.  Specifically we calculate the conditional momentum and conditional position of the mechanical oscillator. Assuming that the mechanical resonator starts in the ground state, the conditional mean amplitude given a photon count at time $t$ is 
\begin{equation}
\langle \Phi^{(1)}(t)|b|\Phi^{(1)}(t)\rangle =\left [R_1(t)\right ]^{-1} \kappa\langle \phi_1(t)|b|\phi_1(t)\rangle
\end{equation}
where the normalisation is given by the single photon count rate in Eq.(\ref{detection_rate}). Using the result in Eq.(\ref{conditional-mech1}), we can write this as 
\begin{eqnarray}
\langle \Phi^{(1)}(t)|b|\Phi^{(1)}(t)\rangle & = & \left [R_1(t)\right ]^{-1}\left (\frac{G_0}{\omega_m}\right )\langle \beta|\hat{R}^\dagger(t) \hat{R}(t)|\beta\rangle+\langle \beta|\hat{R}^\dagger(t) b\hat{R}(t)|\beta\rangle \\
& = & \left [R_1(t)\right ]^{-1}\left (\frac{G_0}{\omega_m}\right )\sum_{n=0}^\infty  e^{-|\beta|^2}\frac{|\beta|^{2n}}{n!} r_n^*(t)[r_n(t)-r_{n+1}(t)]
\label{moment}
\end{eqnarray}
where $r_n(t)$ is given in Eq.(\ref{diagonal_R}). Hence from Eq.(\ref{moment}), we can calculate the conditional moment, $|\langle b \rangle|$ and thus the amplitude of the conditional momentum, $|i\langle (b^{\dagger}-b) \rangle|$. A long detection time $t=t_{1}$, allows the photon to interact longer with the mechanical mode leading to an enhanced OM cooperativity which results in a displaced conditional mechanical state.  

\begin{figure}[!ht]
\centering
\includegraphics[scale=0.6]{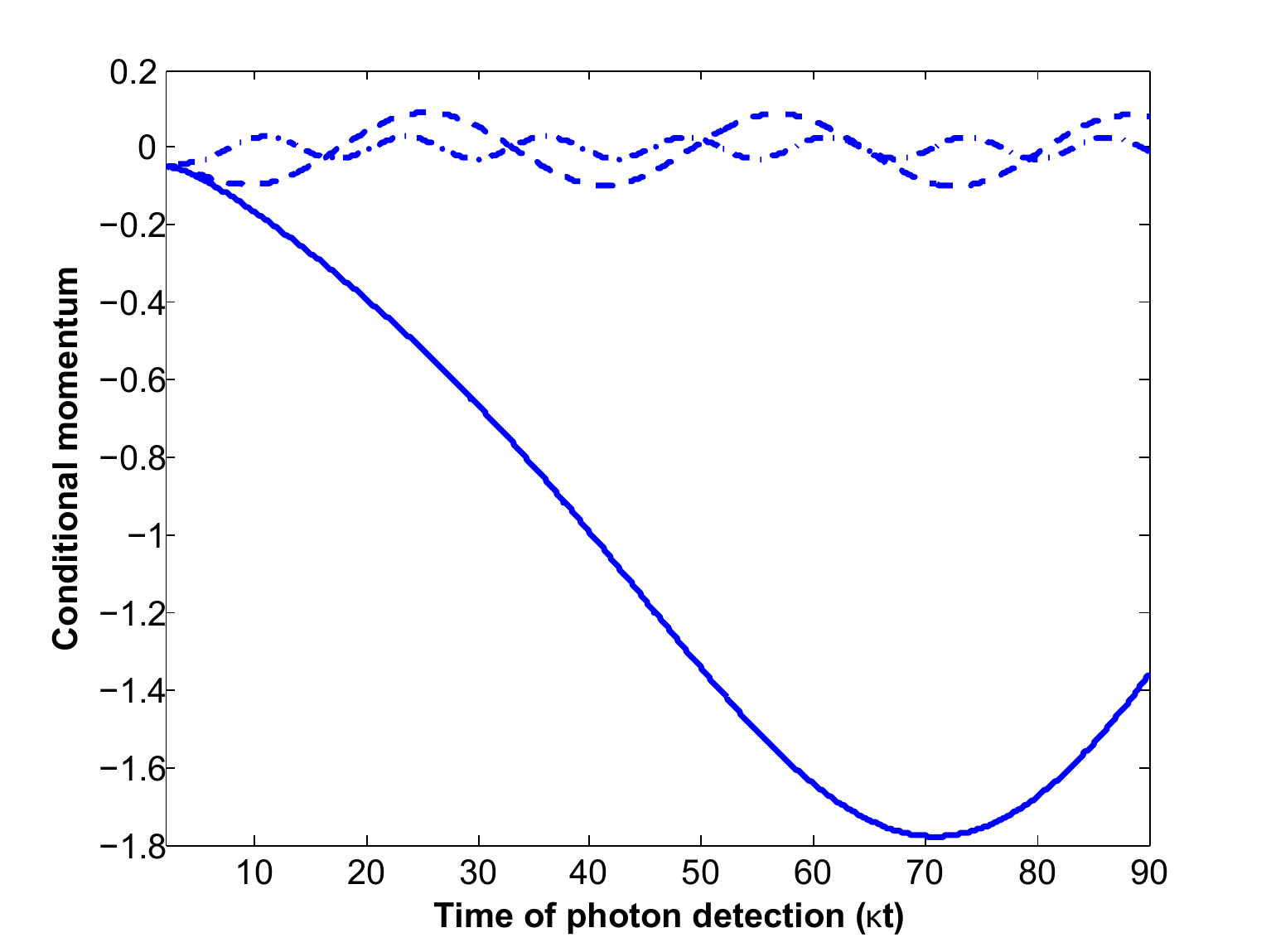}
\caption{The conditional momentum of the mechanical oscillator vs. detection time. Parameters are taken to be dimensionless in units of the optical decay rate $\kappa$: $\gamma/\kappa=2$, $G_{0}/\kappa=0.01$ and different $\omega_{m}/\kappa$: 0.5 (dashed-dotted curve), 0.2 (dashed curve) and 0.02 (solid curve).} 
\label{xp}   
\end{figure}
In Fig.\ref{xp} we have plotted the conditional momentum as a function of time for different values of the mechanical frequency $\omega_{m}$. Initially as the photon enters the OM cavity, it circulates within the OM system, inducing an OM interaction which imparts momentum on the mechanical mode. The maximum amplitude change in the conditional momentum occurs at quarter cycle, $t=T_{m}/4$. This is reflected in Fig.\ref{xp}, where we observe that for mechanical resonators with smaller values of mechanical frequency, postselecting on late detection times of the interacting photon will impart a large momentum kick to the mechanical oscillator. In fact from Eq.(\ref{h_om}) in the semi-classical limit the momentum can be approximated as $-i\langle b-b^{\dagger}\rangle=\frac{-2G_{0}}{\omega_{m}}{\rm sin}(\omega_{m}t)$. In the limit that $\omega_{m}t\ll 1$, where we are considering times short compared to the mechanical period but long compared to the cavity decay rate, $-i\langle b-b^{\dagger}\rangle\thickapprox -2G_{0}t$.

Hence if we focus on the non-resolved sideband regime, $\kappa\gg\omega_{m}$ there is a linear relation between the conditional momentum of the mechanical oscillator with the photon detection time.

\section{Conditional single photon interferometry}
\label{MZ}

In this section we illustrate how single photon conditional optomechanics can be used to generate macroscopic superposition states. 
\begin{figure}[h!]
\centering
\includegraphics[scale=1.3]{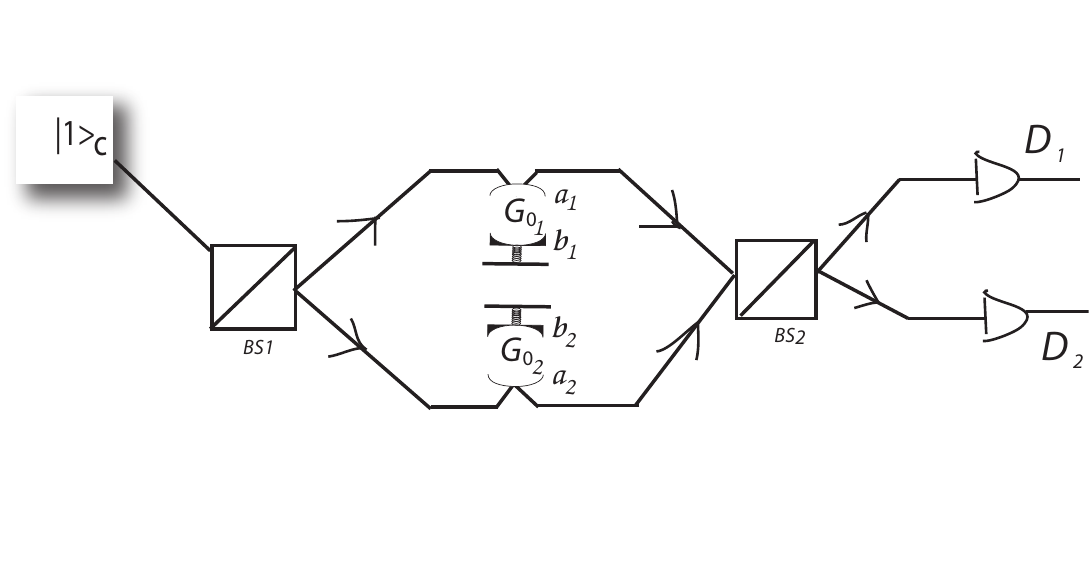}
\caption{Two OM cavities set up as a Mach-Zhender interferometer with detection ports $D_{1}$ and $D_{2}$. The beam splitters $BS_{1,2}$ are considered to have 50:50 transmission and reflection.} 
\label{interferometer}   
\end{figure}

We probabilistically condition two OM cavities set-up as in a Mach-Zhender interferometer, shown in Fig.\ref{interferometer}, with a single driving photon. As previously, the photon source is modelled as an independent cavity coupled irreversibly via the cascaded systems approach to the two OM systems. The arriving photon at the first beam splitter, $BS_1$ can be diverted to either OM system, where it interacts with the mechanical oscillator $b_{i}$. After interaction, the output from the OM cavity again is split at a second beam splitter, $BS_{2}$ before being conditioned on no detections for a time $t$ and then detected at a time $t+dt$ at one of the detection ports, $D_{i}$. Both OM cavities are taken to have identical OM coupling strengths, and the same optical and mechanical resonance frequencies. 

The interaction Hamiltonian for the composite system is an extension of Eq.(\ref{h_om}), given as
\begin{equation}
\label{om-ham2}
H_{om}=\sum_{i=1}^{2}{\hbar\omega_{m_{i}} b_{i}^\dagger b_{i}+\hbar G_{0_{i}} a_{i}^\dagger a_{i} (b_{i}+b_{i}^{\dagger})}
\end{equation}
The master equation and the total Hamiltonian will follow as in Eqs.(\ref{ME}) and (\ref{Htot}), with the cascaded coupling Hamiltonian for each OM cavity given as, 
\begin{equation}
H_{cas} =-i\sqrt{\kappa\gamma}(c a^\dagger-c^\dagger a)/2\sqrt{2}
\end{equation}
where the factor $1/\sqrt{2}$ is due to the effect of the first beam splitter on the arriving photon, such that it has equal probability of entering either arm of the interferometer. After the conditioned interaction with either OM cavity, the output photon goes through the second beam splitter again with equal probability of being diverted to either detection port. Consequently the jump operators for each detection port are given as
\begin{eqnarray}
J_{D_{1}}&=&\sqrt{\gamma}c+\frac{\sqrt{\kappa_{1}}a_{1}+\sqrt{\kappa_{2}}a_{2}}{\sqrt{2}} \\
J_{D_{2}}&=&\frac{\sqrt{\kappa_{1}}a_{1}-\sqrt{\kappa_{2}}a_{2}}{\sqrt{2}}
\end{eqnarray}
where $\kappa_{i}$ is the decay rate of each optical resonator, and the phase differences across the branches of the two beam splitters are reflected as a null detection rate at the port $D_{2}$. The no-jump dynamics of the system is governed similar to Eqs.(\ref{psi}), which in this case is
\begin{equation}
\label{psi2}
|\tilde{\Psi}^{(0)}(t)\rangle =\exp\left [-\frac{i}{\hbar}Ht-\frac{1}{2} J_{D_{1}}^\dagger J_{D_{1}}t- \frac{1}{2} J_{D_{2}}^\dagger J_{D_{2}}t\right ]|\tilde{\Psi}(0)\rangle,
\end{equation}
and evolves as
\begin{equation}
\label{non-hermitian2}
\frac{d |\tilde{\Psi}^{(0)}(t)\rangle}{dt} = -i\left (\frac{H}{\hbar}-\frac{i}{2} J_{D_{1}}^\dagger J_{D_{1}}-\frac{i}{2} J_{D_{2}}^\dagger J_{D_{2}}\right )|\tilde{\Psi}^{(0)}(t)\rangle  
\end{equation}
 Here the basis of the system is defined by the three different cavities in which the photon could exist before being counted, i.e. the source cavity, OM cavity 1 or OM cavity 2; such that at $t=0$, the initial state of the composite system is,
\begin{equation}
\label{ansatz2}
|\tilde{\Psi}(0)\rangle =\left (|1\rangle_{c}|0\rangle_{a_{1}}|0\rangle_{a_{2}}\right )|0\rangle_{b_{1}}|0\rangle_{b_{2}}
\end{equation}
where we have taken each mechanical oscillator to start in the ground state at $t=0$. 

Transforming to the displacement picture we proceed as before such that a photon count at $D_{1}$ gives the unnormalised conditional state of the composite mechanical system as
\begin{equation}
|\tilde{\Phi}^{D_{1}}(t)\rangle = \left(-\frac{\kappa_{1}}{2} D^{\dagger}_{1}(\beta_{1})\hat{R}_{1}D_{1}(\beta_{1})-\frac{\kappa_{2}}{2} D^{\dagger}_{2}(\beta_{2})\hat{R}_{2}D_{2}(\beta_{2})+1\right)|0\rangle_{b_{1}}|0\rangle_{b_{2}}\\
\label{cat}
\end{equation}
where the $\hat{R}_{i}$ are each given as in Eq.(\ref{hatR}), and $\beta_{i}=G_{0_{i}}/\omega_{m_{i}}$. 

It can immediately be seen that the conditional state in Eq.(\ref{cat}) has approximately the form of an entangled cat state
\begin{equation}
|\psi(t)\rangle=a_{1}(t)|\phi_{1}\rangle_{1}|0\rangle_{2}+a_{2}(t)|0\rangle_{1}|\phi_{2}\rangle_{2}+a_{3}(t)|0\rangle_{1}|0\rangle_{2}
\label{sample_state}
\end{equation}
where the amplitude of each mechanical mode is given by $\phi_{i}$, defined in Eq.(\ref{conditional-mech1}). For a given late detection at time $t=t_{1}$ conditioned on no counts up to $t_{1}$, the contribution from the third component of the normalised state, which accounts for a direct detection of the photon from the source, approaches zero.  
We now investigate the effect of postselection on the entanglement and nonclassicality between the mechanical oscillators $b_{1}$ and $b_{2}$. To quantize the entanglement we calculate the Von Neumann entropy, after tracing over one of the mechanical modes. The Von Neumann entropy for a system described by a density matrix, $\rho$ is defined as 
\begin{equation}
\cal{E}=-\rm{Tr}(\rho \rm{ln} \rho)
\end{equation} 
The presence of bipartite entanglement between the resonators is confirmed by nonzero values of the entropy of the reduced state of either of them. 

\begin{figure}[h!]
\centering
\includegraphics[scale=0.5]{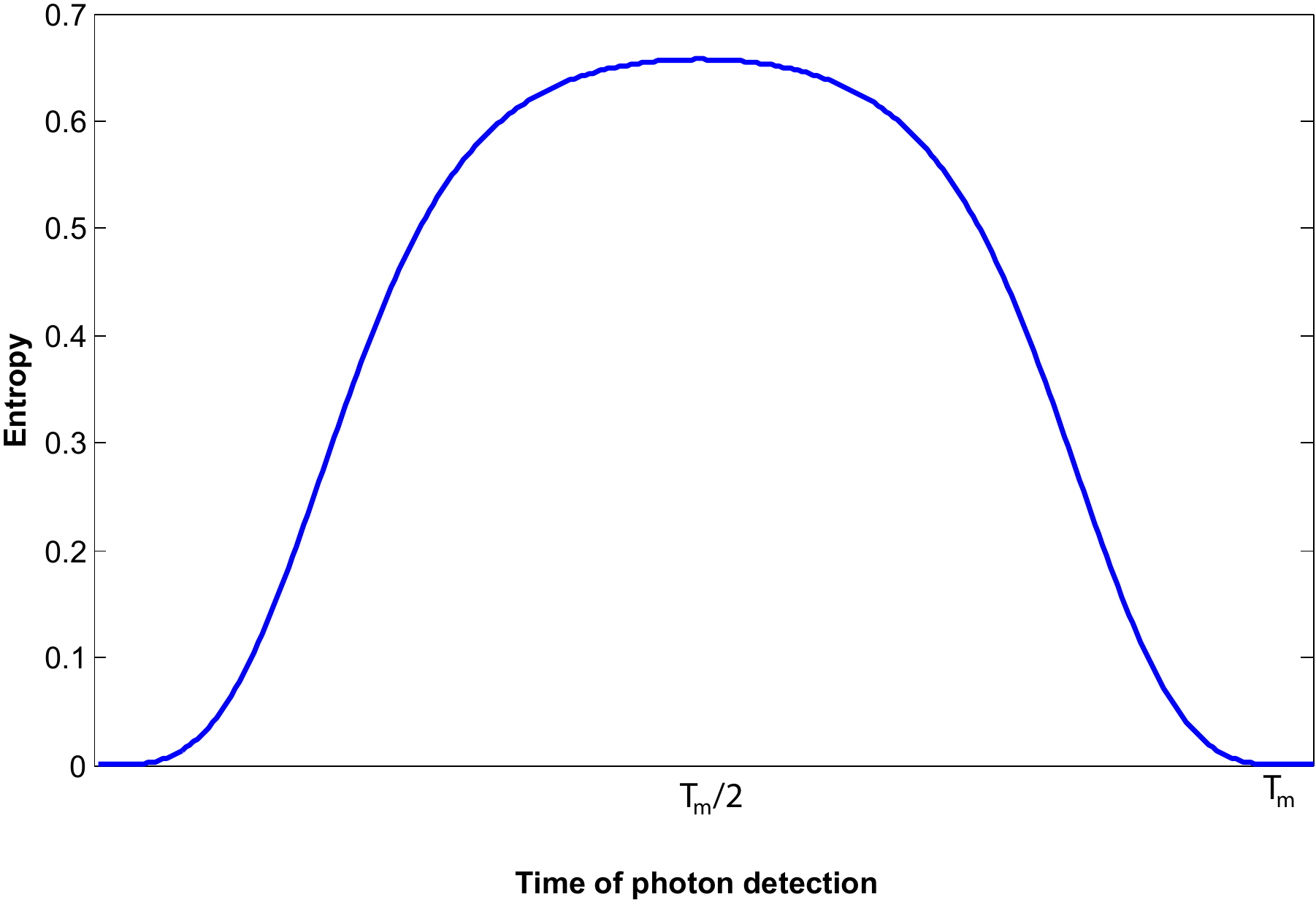}
\caption{The entropy after tracing out one of the mechanical modes of the combined conditional state. Parameters are taken to be dimensionless in units of the optical decay rate $\kappa$ which is taken to be equal for both OM cavities: $\gamma/\kappa=2$, $G_{0_{i}}/\kappa_{i}=0.02$ and $\omega_{m_{i}}/\kappa_{i}=0.02$ vs. time of photon detection within one mechanical period.} 
\label{entropy}   
\end{figure} 

Fig.(\ref{entropy}) shows the Von Neumann entropy, as a function of the duration of conditioning by the driving photon over one full mechanical period for the parameters shown. For the first half-cycle we find the entropy increases to a maximum value before decreasing again in the next half cycle. Hence at half cycle of the mechanical oscillator, when its amplitude in phase space is maximal, the mechanical oscillators $b_{1}$ and $b_{2}$ are maximally entangled.  

In order to probe how the conditional state, given in Eq.(\ref{cat}) evolves during one complete mechanical period, it is useful to analyse the mean amplitude in phase space experienced by each mechanical mode $\langle b_{i} \rangle$ as a result of its interaction with the single photon. 

\begin{figure}[h!]
\centering
\includegraphics[scale=0.7]{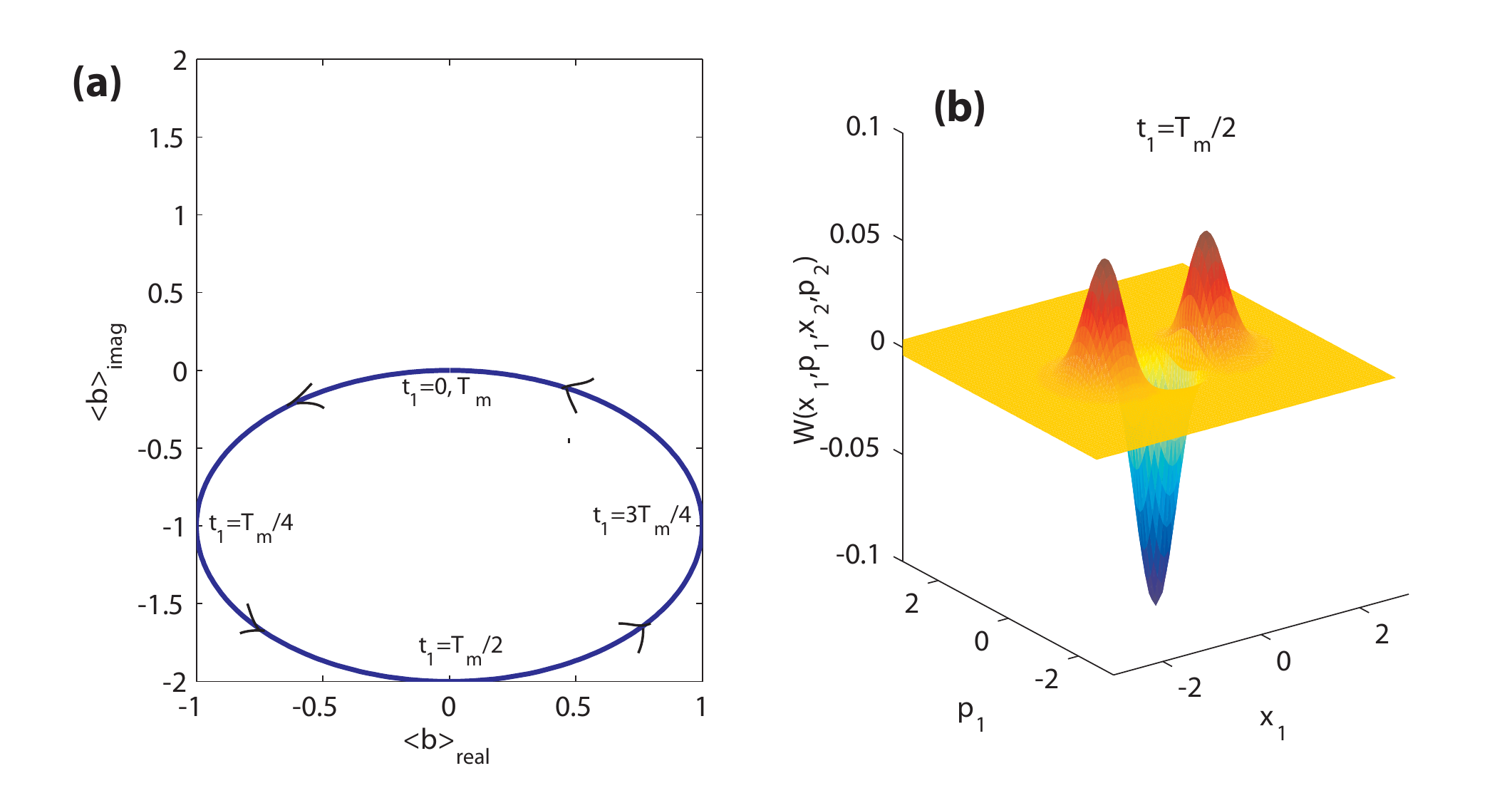}
\caption{(a) A parametric plot of the real and imaginary components of the mean amplitude in phase space for a mechanical mode. (b) A slice of the Wigner function of the combined conditional state of two identical mechanical modes vs. $p_{1}$ and $x_{1}$, projected over a specific $x_{2},p_{2}$ (see text below). The state is detected at half cycle of the mechanical period. Parameters are chosen in units of optical decay rate $\kappa_{i}=\kappa$: $G_{0_{i}}/\kappa=0.02, \omega_{m_{i}}/\kappa=0.02$, $\gamma=2/\kappa$.} 
\label{wigner_and_b}   
\end{figure} 

In Fig.\ref{wigner_and_b}(a), we plot the real and imaginary parts of the mean amplitude in phase space, $\langle b\rangle$ for a mechanical mode which has been driven by a single photon for one complete mechanical period, $T_{m}$. After the mechanical mode has interacted with the photon for a quarter cycle $t_{1}=T_{m}/4$, it is displaced such that its imaginary component is zero, and the real component is maximum. It goes through a phase change after $t_{1}=T_{m}/4$, such that at half cycle of the mechanical period, $t_{1}=T_{m}/2$, the imaginary part of $\langle b \rangle$ is now maximum whereas the real part is zero. At this stage, the mechanical mode experiences the maximum  allowed amplitude in phase space for the parameters chosen. Hence if detected at this specific time, one would expect it to exhibit its optimal nonclassical behaviour as an entangled cat state. For detection times $t\gg T_{m}/2$, the real and imaginary parts of the mean amplitude of the mechanical mode experience a phase shift again and map back their path for the remaining cycle.    

Therefore in order to probe the nonclassicality of the conditional state, we focus on its Wigner function at half cycle. The total Wigner function of such a product state depends on four dimensions: the position $x_{1,2}$ and momentum $p_{1,2}$ of each oscillator. At this stage, we are in the regime of a late detection $t_{1}\gg 0$, and therefore the contribution from the third component of the normalised conditional state given in Eq.(\ref{cat}) is negligible. Hence, we can compare our conditional state at half cycle to a cat state of the form 
\begin{equation}
|\alpha\rangle_{1}|0\rangle_{2} +|0\rangle_{1}|\alpha\rangle_{2}
\label{sample_cat}
\end{equation} 
where $| \alpha\rangle$ is a perfect coherent state. To choose the projection that would give us optimal nonclassical behaviour, we analyse the cat state in Eq.(\ref{sample_cat}) while taking the coherent amplitude as the mean amplitude in phase space of the mechanical oscillator driven by a single photon, i.e. $\alpha=\langle b \rangle_{imag}$ at $t_{1}=T_{m}/2$. We then determine that the particular combination of the axes $x_{1},p_{1},x_{2}=-x_{1},p_{2}=\frac{\pi}{2\alpha}+p_{1}$ results in the largest negativity from the interference term of the combined Wigner function. Taking this specific projection, we plot a slice of the Wigner function of our conditional state, in Fig.\ref{wigner_and_b}(b). As can be seen, our conditional state displays two peaks, with an interference between them resulting in negativity in the Wigner function, as is characteristic of a Schrodinger cat state. Each peak corresponds to the conditional mean amplitude in phase space at half cycle of each mechanical resonator as a result of single photon driving. In the results presented here, a large nonlinearity ($G_{0}/\omega_{m}=1$) has been used, consistent with those possible, for example, in OM systems consisting of an atomic ensemble within an optical cavity \cite{DSK}. For systems with smaller ratios, similar results are achievable by adjusting to a longer detection time thereby increasing the resultant displacement of the conditional mean amplitude of the mechanical oscillator.  

Thus far we have considered both OM cavities with identical mechanical frequencies, $\omega_{m}/\kappa$ and OM coupling strengths, $G_{0}/\kappa$ in each arm of the Mach-Zhender interferometer, for simplicity. 

\begin{figure}[h!]
\centering
\includegraphics[scale=0.5]{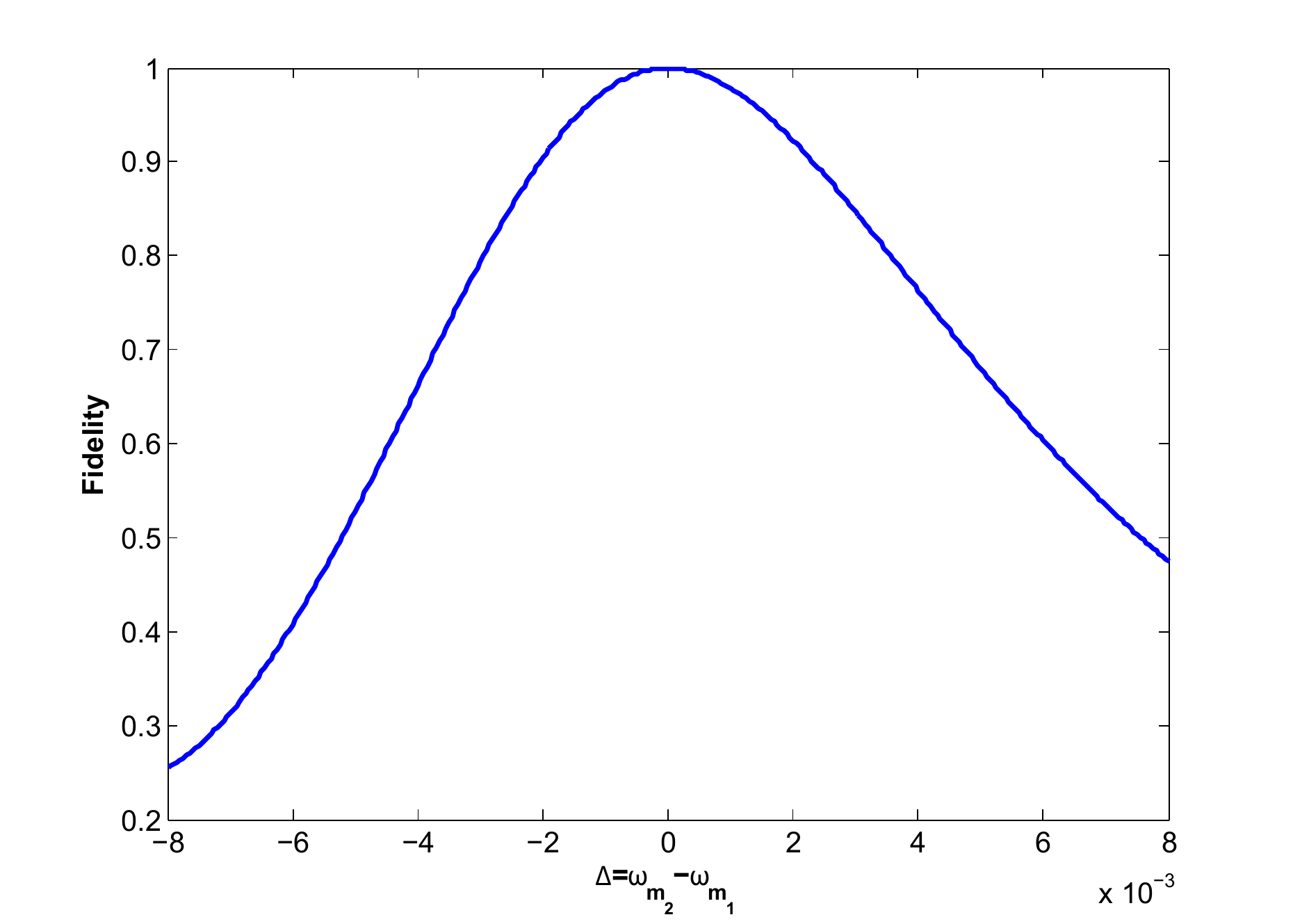}
\caption{Fidelity for obtaining a cat state with identical mechanical frequencies vs detuning $\Delta$ between the mechanical frequencies of the resonators. Parameters are chosen in units of the optical decay rate $\kappa_{i}=\kappa$: $G_{0_{1}}/\kappa=0.02, \omega_{m_{1}}/\kappa=0.02$ and $\gamma/\kappa=2$.}
\label{fid}
\end{figure}
To test if the protocol allows for some detuning between the mechanical frequencies in the two arms, we calculate the fidelity between the conditioned state at half cycle for identical mechanical frequencies with the state for some variation between the frequencies. In Fig.\ref{fid} we plot the fidelity vs. detuning between the mechanical frequencies. For a detuning of $|\Delta|=10\%$, the combined conditional state has a high fidelity with a minimum at $90\%$. For $|\Delta|=20\%$, the conditional state exhibits good fidelity with a minimum of $80\%$.  However for $|\Delta|>20\%$, the fidelity of the conditioned state falls rapidly. Hence our scheme allows some variation between the frequencies of the mechanical oscillators to obtain a cat state with high fidelity measures. 

\section{Two photon conditional dynamics}
\label{2photon}

\begin{figure}[h!]
    \centering
    \includegraphics[scale=0.7]{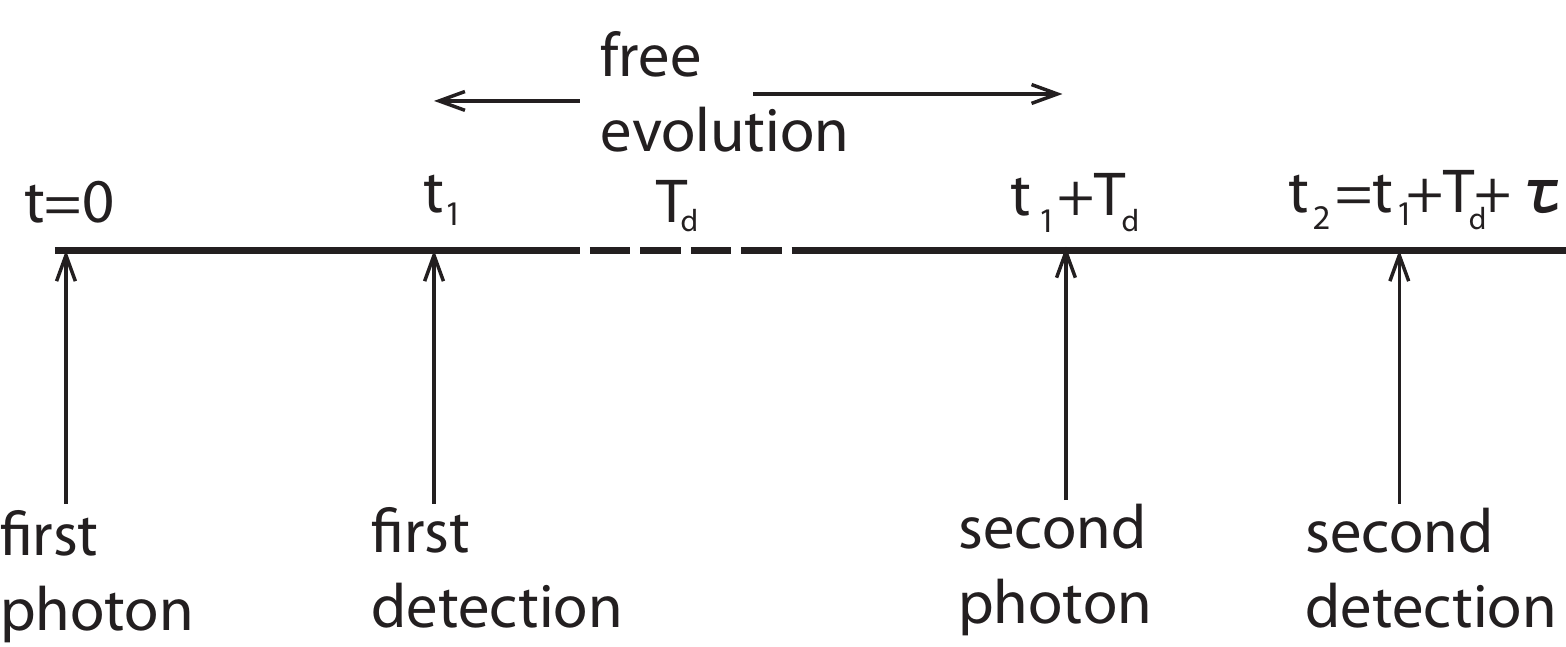} 
    \caption{The temporal protocol for exciting an OM cavity with two consecutive single photon pulses. The first photon prepared in the source cavity at time $t=0$, interacts with the mechanics and is detected at time $t_1$. The mechanical resonator then evolves freely for a time delay, $T_{d}$, between detection of the first photon at $t_1$ and preparation of the second photon. The second photon then arrives at $t_{1}+T_{d}$ interacts for a time $\tau$ before being detected at time $t_2=t_1+T_{d}+\tau$.}
    \label{fig_protocol}
 \end{figure}

 We now turn towards probing further the characteristics of a conditioned OM system, by considering the two-photon excitation protocol shown in Fig.\ref{fig_protocol}.
  We wish to compute the rate of detection of the second photon as a function of the delay time, $T_{d}$ and duration of its interaction with the mechanical oscillator, $\tau$.  The conditional (normalised) state of the mechanical system given that the first photon was counted at time $t_1$ in Eq.(\ref{conditional-count}) can be written in the form 
  \begin{equation}
  |\Phi^{(1)}(t_1)\rangle =\sqrt{\kappa}|T\rangle+\sqrt{\gamma}|R\rangle
  \end{equation}
  which is a superposition of two histories: detection after transmission through the cavity, $|T\rangle$ and detection after reflection from the cavity $|R\rangle$. 
  
We now allow this conditional state to evolve freely for a time of duration $T_{d}$, while the source is prepared with another single photon. Given that this second photon interacts for a time $\tau$ with the OM cavity and is then detected at a time $t_2=t_1+T_{d}+\tau$, the final conditional state the system evolves to is a result of applying the jump operator twice to the initial state at $t_{1}=0$. There are now four indistinguishable temporal histories for the second detection so that the conditional (unnormalised) state of the mechanical resonator is given by
\begin{equation}
|\Phi^{(2)}(t_2:T_{d}:t_1:0)\rangle = \kappa |TT\rangle+\gamma|RR\rangle+\sqrt{\kappa\gamma}(|RT\rangle+|TR\rangle)
\label{phi2}
\end{equation}

The {\em conditional} rate of detection of the second photon, $R_{2}$ can be evaluated as 
\begin{eqnarray}
\label{ratephoton2}
R_{2}(t_{2},T_{d},t_{1})&=&\langle \phi^{(2)}(t_2:T_{d}:t_1:0)|\phi^{(2)}(t_2:T_{d}:t_1:0)\rangle\\ \nonumber
& & =  \kappa^2\langle TT|TT\rangle +\gamma^2\langle RR|RR\rangle\\ \nonumber
& & +\kappa\gamma\left [ \langle RT|RT\rangle + \langle TR|TR\rangle\right ]\\ \nonumber
& &+\kappa\gamma\left [ \langle TT|RR\rangle  +\langle RT|TR\rangle+c.c.)\right ]\\ \nonumber
& & +\kappa\sqrt{\kappa\gamma}\left [\langle TT|RT\rangle+\langle TT|TR\rangle+c.c.\right ]\\ \nonumber
& & +\gamma\sqrt{\kappa\gamma}\left [\langle RR|RT\rangle+\langle RR|TR\rangle +c.c.\right ]
\end{eqnarray}

Explicit expressions for each doubly conditioned part $|XY\rangle$ of the state $|\Phi^{(2)}\rangle$ as well as for each term of Eq.(\ref{ratephoton2}) are given in the Appendix. 
   
 The conditional rate of detection of the second photon, $R_{2}$ can be plotted vs. the free evolution time, $T_{d}$ and interaction time of the second photon, $\tau=t_{2}-t_{1}-T_{d}$ for given values of detection times of the first photon, $t_{1}$. Similar to the one photon case, the detection rate is composed of the rate of reflected and transmitted photons as well as the interference term arising from transmitted and reflected photons. 

\begin{figure}[h!]
\centering
\includegraphics[scale=0.6]{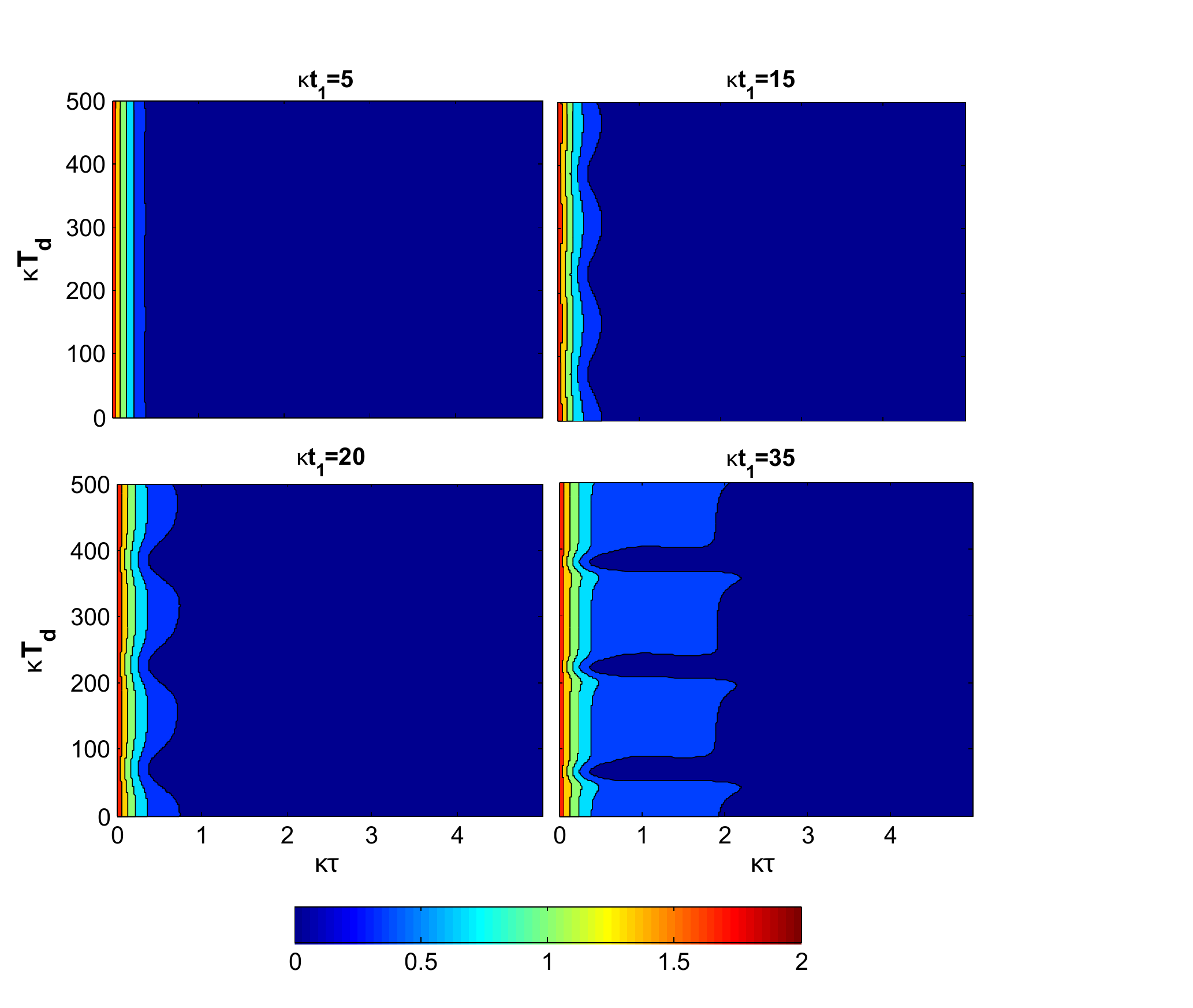}
\caption{The total conditional rate of detection of the second photon as a function of its OM interaction time $\kappa\tau$ and free evolution $\kappa T_{d}$ for $\gamma/\kappa=2$, $G_{0}/\kappa=0.05$, $\omega_{m}/\kappa=0.02$ and different detection times of the first photon $\kappa t_{1}$.} 
\label{rate2}   
\end{figure} 

Fig.~\ref{rate2} shows the total conditional rate of detection for the second photon. Figures showing the effect of late detections of the first photon on the individual reflected, transmitted and interference parts of the conditional rate of detection of the second photon can be found in the Appendix.  In Fig.\ref{rate2} we find for early detection times of the first photon, the profile for the total conditional rate of the second photon remains unchanged on the free evolution axis. However if the first photon is conditioned on late detections, we observe a periodic detuning effect occurring along the free evolution axis which becomes sharper as the interaction time of the first photon increases. This is a consequence of the larger displacement and hence momentum imparted to the mechanical oscillator. The displacement of the mechanical oscillator is accompanied by a change in the frequency of the OM cavity, taking it off resonance from the single photon source it is coupled to. Consequently, at quarter cycle after arrival of the first photon, the detuning between the OM cavity and the single photon source will be maximum. At this point in time along the free evolution axis, the subsequent second photon is unable to couple to the OM cavity and is thus routed off.
\begin{figure}[h!]
\centering
\includegraphics[scale=0.8]{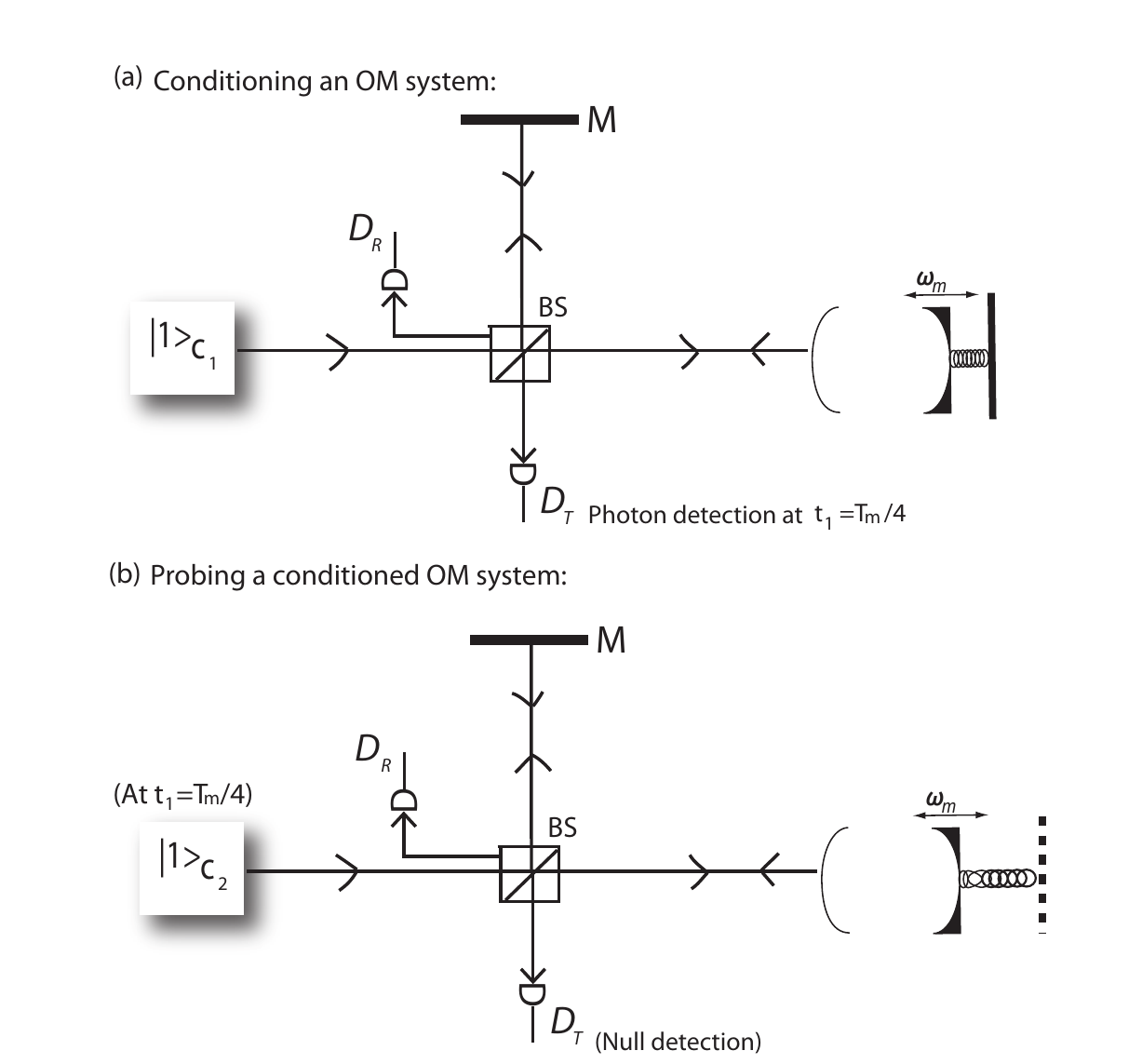}
\caption{{\em Engineering a single photon router using two photon conditional optomechanics in a Michelson interferometer:} Detectors $D_{R}$ and $D_{T}$ respectively condition only reflected and transmitted possibilities. (a) Detection of a photon $|1\rangle_{c_{1}}$, by detector $D_T$ at time $t_1=T_{m}/4$ conditions the mechanical oscillator with a well defined momentum kick. (b) A second photon, $|1\rangle_{c_{2}}$, interacting with the OM system at $t_1=T_m/4$ will find the optical cavity off resonance, and will be routed to detector $D_R$.}
\label{router}   
\end{figure}

 This effect thereafter occurs every half-cycle. Such a doubly-conditioned cavity can be employed as a periodic single photon router for subsequent photons, given late photon counts for the initial photon as illustrated in Fig.\ref{router}. Again a suitable candidate to test these effects could be the low frequency, SiN string mechanical element with large Q described in \cite{schmid}. 

\section{Summary}
\label{summary}
We have analysed in detail single photon optomechanics conditioned on photon counting events in this work. Our results show how late photon counts in an OM cavity can enhance cooperativity between the optical and mechanical modes such that a large momentum can be imparted to the mechanical mirror. Applying this idea to the interaction of a single photon with a pair of OM cavities arranged as in an interferometer, we calculate a combined conditional state of the two mechanical modes in the composite system. Such a conditional state can become an entangled mechanical cat state corresponding to photon counting events close to half a mechanical period. Thus this paper quantifies the idea of generating macroscopic superposition states using conditional single photon driving of two OM systems. Further we have also described a two photon conditioned OM protocol. We show how injecting a single photon into an already conditioned cavity after a quarter of a mechanical period can allow the OM system to act as a periodic single photon router which could have varied applications in quantum information networks. 

These results may be realised using SiN string mechanical elements, which can be cooled to their quantum ground state, or prepared in a coherent state via the recently proposed protocols in the pulsed optomechanics regime \cite{vanner1}. Another suitable experimental avenue to implement the findings in this paper are hybrid OM systems where the mechanical element is an ensemble of cold atoms trapped in an optical field \cite{DSK}. Our choice of parameters match well with recent parameters reported for these systems which are also in the nonresolved sideband regime. Crucially they have low, tunable mechanical frequencies and $G_{0}/\omega_{m}$ as high as $0.4$ has been reported. Finally we emphasise that since the proposals introduced here involve single photon pulses and single photon conditioning, once a detection event occurs the total efficiency of the system is conditioned to 100\%. Consequently, inefficiency is not a crucial concern in these proposals, with its only effect being to change the rate at which successful experiments occur and therefore the overall time duration required to gain useful statistics from experiments. 

\section*{Acknowledgements}
We wish to acknowledge the support of the Australian Research Council CE110001013 through the Centre of Excellence for Engineered Quantum Systems. UA also acknowledges support from the University of Queensland Postdoctoral Research Fellowship and Grant. 

 \section{Appendix}
 The relevant terms in Eq.(\ref{phi2}) are given as
\begin{eqnarray}
|RR\rangle & = &  e^{-i\omega_mb^\dagger b t_2-\gamma(t_2-T_{d})/2}|\psi(0)\rangle\\
|RT\rangle & = &  e^{-i\omega_m b^\dagger b(t_2-t_1)-\gamma\tau/2}\hat{K}(t_1:0)|\psi(0)\rangle\\
|TR\rangle & = &  \hat{K}(t_2:t_1+T_{d})e^{-i\omega_m b^\dagger b (t_1+T_{d})-\gamma t_1/2}|\psi(0)\rangle\\
|TT\rangle & = &  \hat{K}(t_2:t_1+T_{d})e^{-i\omega_m b^\dagger b T_{d}}\hat{K}(t_1;0)|\psi(0)\rangle
\end{eqnarray}
    
Each of the required terms in Eq.(\ref{ratephoton2}) can be evaluated as,
  \begin{eqnarray}
  \langle TT|TT\rangle & = & \langle 0|\hat{K}^\dagger(t_1:0) e^{i\omega_m b^\dagger b T_{d}}\hat{K}^\dagger(t_2;t_1+T_{d})\hat{K}(t_2;t_1+T_{d})e^{-i\omega_m b^\dagger b T_{d}}\hat{K}(t_1;0)|0\rangle\\ \nonumber
  \langle RR|RR\rangle & = & e^{-\gamma(t_2-T_{d})}\\ \nonumber
  \langle RT|RT\rangle & = & e^{-\gamma\tau}\langle 0|\hat{K}^\dagger(t_1;0)\hat{K}(t_1;0)|0\rangle \\ \nonumber
  \langle TR|TR\rangle  & = & e^{-\gamma t_1}\langle 0|\hat{K}^\dagger(t_2;t_1+T)\hat{K}(t_2;t_1+T_{d}) |0\rangle  \\ \nonumber
  \langle TT|RR\rangle   & = & e^{-\gamma(t_2-T_{d})/2}\langle 0|\hat{K}^\dagger(t_1;0) e^{i\omega_m b^\dagger b T_{d}}\hat{K}^\dagger(t_2;t_1+T_{d}) |0\rangle\\ \nonumber
  \langle RT|TR\rangle & = & e^{-\gamma(t_2-T_{d})}\langle 0|\hat{K}^\dagger(t_1;0) e^{i\omega_m b^\dagger b (t_2-t_1)}\hat{K}(t_2;t_1+T_{d}) |0\rangle\\ \nonumber
  \langle TT|RT\rangle & = & e^{-\gamma \tau/2}\langle 0|\hat{K}^\dagger(t_1;0) e^{i\omega_m b^\dagger bT_{d}}\hat{K}^\dagger(t_2;t_1)e^{-i\omega_m b^\dagger b (t_2-t_1)}\hat{K}(t_1;0) |0\rangle\\ \nonumber
  \langle TT|TR\rangle & = & e^{-\gamma t_1/2}\langle 0|\hat{K}^\dagger(t_1;0) e^{i\omega_m b^\dagger bT_{d}}\hat{K}^\dagger(t_2;t_1+T_{d}) \hat{K}(t_2;t_1+T_{d})|0\rangle\\ \nonumber
  \langle RR|RT\rangle  & = & e^{-\gamma(t_1+2\tau)/2}\langle 0|\hat{K}(t_1;0)|0\rangle\\ \nonumber
  \langle RR|TR\rangle & = & e^{-\gamma(t_2-T_{d})/2}\langle 0|\hat{K}(t_2;t_1+T_{d})e^{-i\omega_m b^\dagger bT_{d}}\hat{K}(t_1;0)|0\rangle.
  \end{eqnarray}

  \begin{figure}[h!] 
\centering
\includegraphics[scale=0.6]{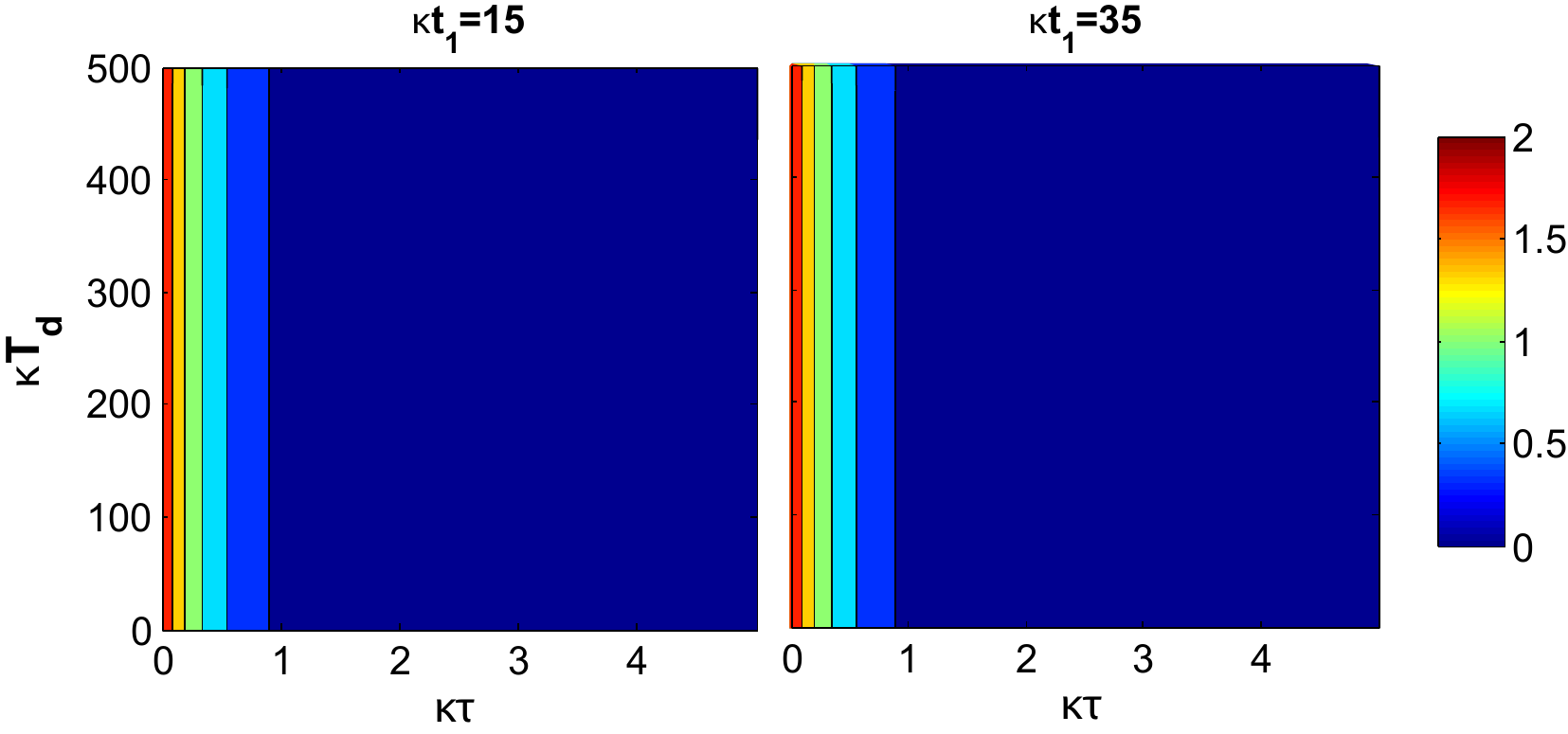}
\caption{The reflected part of the conditional rate of detection of the second photon as a function of its OM interaction time $\kappa \tau$ and free evolution $\kappa T_{d}$ for $\gamma/\kappa=2$, $G_{0}/\kappa=0.1$, $\omega_{m}/\kappa=0.02$ and different $\kappa t_{1}$.} 
\label{reflect2}   
\end{figure} 

 Fig.~\ref{reflect2} shows the rate of detection of the second photon corresponding to reflected photons. This part of the conditional rate is unaffected by the detection time of the first photon, as expected. 
 
\begin{figure}[h!] 
 \centering
\includegraphics[scale=0.6]{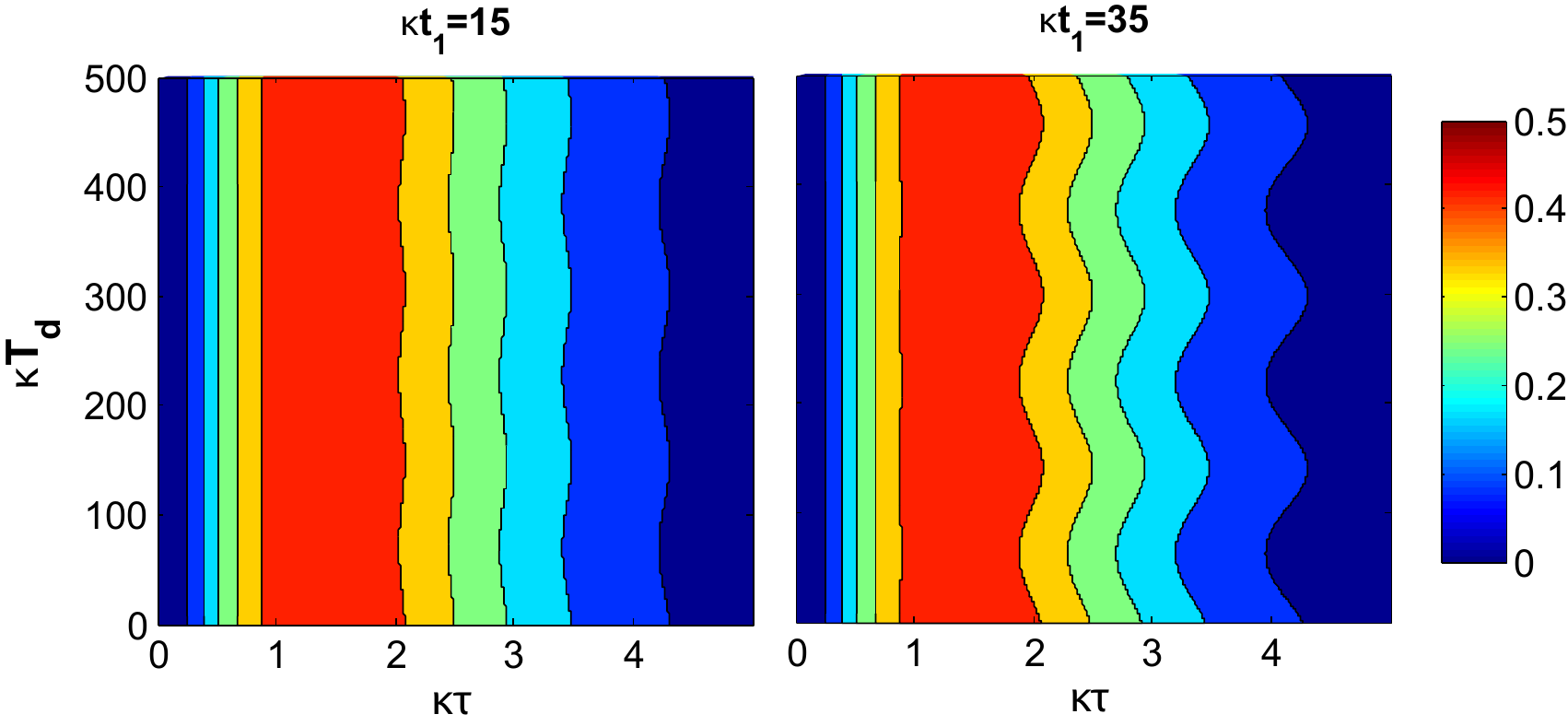}
\caption{The transmitted part of the conditional rate of detection of the second photon as a function of its OM interaction time $\kappa \tau$ and free evolution $\kappa T_{d}$ for $\gamma/\kappa=2$, $G_{0}/\kappa=0.1$, $\omega_{m}/\kappa=0.02$ and different detection times of the first photon $\kappa t_{1}$.} 
\label{trans2}   
\end{figure}  

Fig.~\ref{trans2} shows the part of the conditional rate of detection of the second photon arising specifically from photons which were transmitted into the OM cavity, interacted with the mechanical oscillator before being emitted towards the photo detector. 

\begin{figure}[h!]
\centering
\includegraphics[scale=0.6]{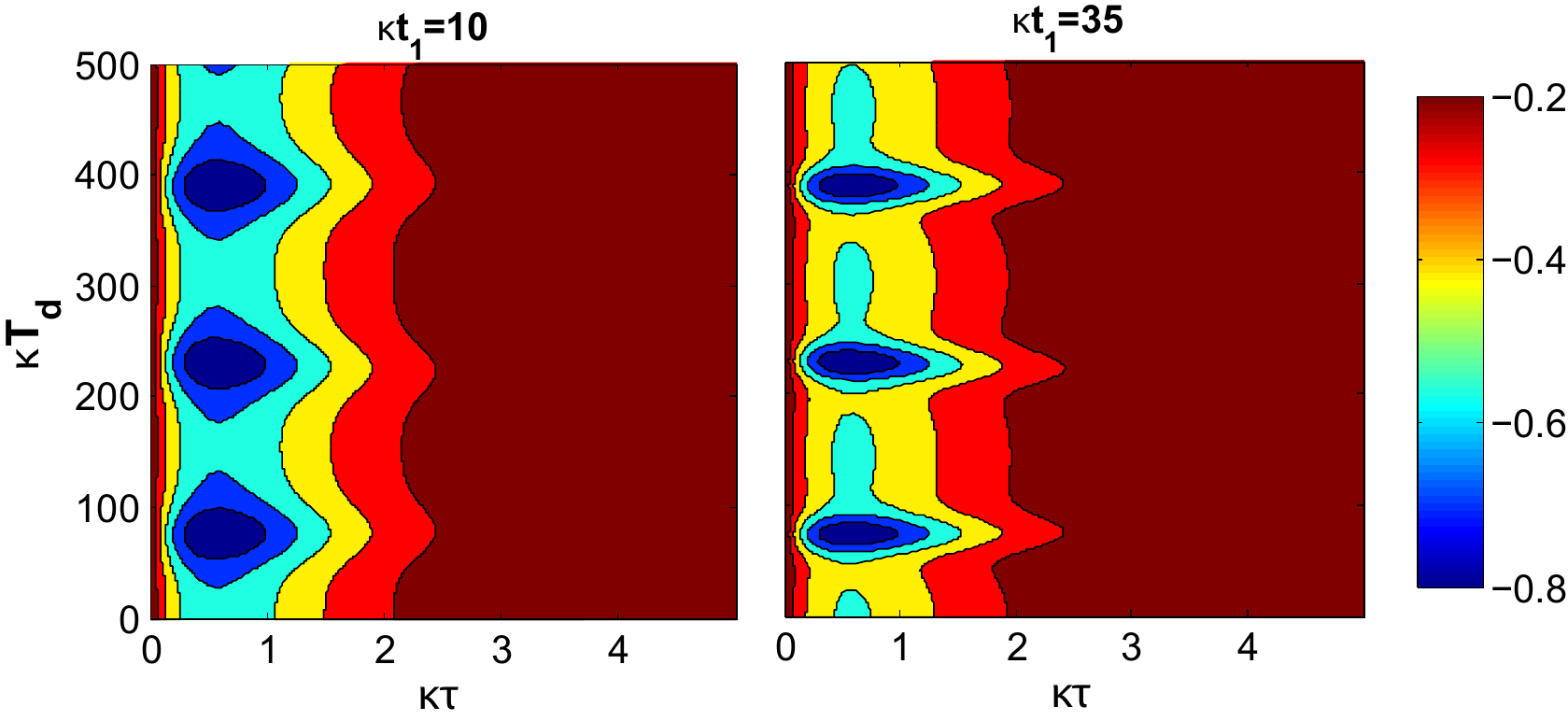}
\caption{The interference part of the conditional rate of detection of the second photon as a function of its OM interaction time $\kappa \tau$ and free evolution $\kappa T_{d}$ for $\gamma/\kappa=2$, $G_{0}/\kappa=0.1$, $\omega_{m}/\kappa=0.02$ and different detection times of the first photon $\kappa t_{1}$.} 
\label{interference2}   
\end{figure} 
The interference part of the conditional rate is shown in Fig.~\ref{interference2} for different times $\kappa t_{1}$ of detection of the first photon. This part of the conditional rate arises from a change of phase and is thus most affected by the prolonged interaction of the first driving photon.


\begin{references}
\bibitem{ybchen} F. Khalili, S. Danilishin, H. Miao, H. Muller-Ebhardt, H. Yang and Y. Chen, Phys. Rev. Lett, {\bf 105} 070403, 2010.
\bibitem{aspelmeyer} S. Gr\"{o}blacher, K. Hammerer, M. R Vanner, M. Aspelmeyer, Nature, {\bf 460} 724 (2009).
\bibitem{akram} U. Akram, N. Kiesel, M. Aspelmeyer and G. J. Milburn, New J. Phys., {\bf 12}, 083030 (2010).
\bibitem{Wilson-Rae} I. Wilson-Rae, N. Nooshi, W. Zwerger,1 and T. J. Kippenberg, Phys. Rev. Lett. {\bf 99}, 093901 (2007) .
\bibitem{vitali} D. Vitali, S. Gigan, A. Ferreira, H. R. Bohm, P. Tombesi, A. Guerreiro, V. Vedral, A. Zeilinger, and M. Aspelmeyer, Phys. Rev. Lett. 98, 030405 (2007).
\bibitem{genes} C. Genes, A. Mari, P. Tombesi, and D. Vitali, Phys. Rev. A 78, 032316 (2008).
\bibitem{gangat} A.A.Gangat, T.M.Stace, G.J.Milburn, New J. Phys. {\bf 13}  043024 (2011).  
\bibitem{Kippenberg}E. Verhagen, S. Deleglise, S. Weis, A. Schliesser, and T. J. Kippenberg,  Nature {\bf 482}, 63, (2012). 
\bibitem{connell}A. D. O'Connell, M. Hofheinz, M. Ansmann, R. C. Bialczak, M. Lenander, E. Lucero, M. Neeley, D. Sank, H. Wang, M. Weides, J. Wenner, J. M. Martinis, A. N. Cleland, Nature, {\bf 464},  697 (2010). 
\bibitem{teufel} J. D. Teufel, D. Li, M. S. Allman, K. Cicak, A. J.  Sirois, J. D. Whittaker,  and R. W. Simmonds, Nature {\bf 471}, 204-208 (2011).
\bibitem{painter} J. Chan, T. P. Mayer Alegre, A. H. Safavi-Naeini, J. T. Hill, A. Krause, S. Groeblacher, M. Aspelmeyer and O. Painter, Nature {\bf 478}, 89-92 (2011). 
\bibitem{xuereb} A. Xuereb, C. Genes and A. Dantan, Phys. Rev. Lett. {\bf 109}, 223601 (2012).
\bibitem{rabl}  P. Rabl, Phys. Rev. Lett, {\bf 107} 063601, (2011).
\bibitem{nunnenkamp}A. Nunnenkamp, K. Borkje and S. M. Girvin, Phys. Rev. Lett. {\bf 107} 063602 (2011). 
\bibitem{bing} Bing He, Phys. Rev A {\bf 85}, 063820 (2012). 
\bibitem{jieo-qiao} Jieo-Qiao Laio, H.K.Cheung and C.K.Law, Phys. Rev. A {\bf 85} 025803 (2012).
\bibitem{xun-wei2} X.W. Xu, Y. J. Li and Yu-xi Liu, Phys. Rev. A {\bf 87}, 025803 (2013)
\bibitem{kronwald} Andreas Kronwald, Max Ludwig and Florian Marquardt Phys. Rev. A {\bf 87}, 013847 (2013).
\bibitem{xun-wei} Xun-Wei Xu and Yuan-Jie Li, J. Phys. B: At. Mol. Opt. Phys. {\bf 46} 035502 (2013).
\bibitem{bose} S. Bose, K. Jacobs and P.L. Knight, Phys Rev. A {\bf 59}, 3204 (1999). 
\bibitem{marshall} W. Marshall, C. Simon, R. Penrose, and D. Bouwmeester,
Phys. Rev. Lett. {\bf 91}, 130401 (2003).
\bibitem{bouwmeester} Brian Pepper, Roohollah Ghobadi, Evan Jeffrey, Christoph Simon and Dirk Bouwmeester,  Phys. Rev. Lett. {\bf 109} 023601 (2012).
\bibitem{ybchen2} Ting Hong, Huan Yang, Haixing Miao and Yanbei Chen, Phys. Rev. A, {\bf 88}, 023812 (2013). 
\bibitem{vanner2} M.R. Vanner, M. Aspelmeyer and M.S. Kim, Phys. Rev. Lett. {\bf 110}, 010504 (2013).
\bibitem{Liao} Jie-Qiao Liao, C. K. Law , Phys. Rev. A {\bf 87}, 043809 (2013).
\bibitem{walmsley} K. C. Lee et. al, Science {\bf 334}, 1253 (2011).
\bibitem{Li} Jie Li, Simon Groblacher and Mauro Paternostro, New Journal of Physics, {\bf 15}, 033023 (2013). 
\bibitem{borkje} K.Borkje, A. Nunnenkamp and S.M.Girvin, Phys. Rev. Lett. {\bf 107}, 123601 (2011). 
\bibitem{schmid} S. Schmid, K.D.Jensen, K.H.Nielsen and A.Boisen, Phys. Rev. B {\bf 84} 165307, (2011).
\bibitem{painter2} M. Ludwig, A. H Safavi-Naeini, O. Painter and F. Marquardt, Phys. Rev. Lett. {\bf 109}, 063601, (2012). 
\bibitem{jharris1} J. D. Thompson, B. M. Zwickl, A. M. Jayich, Florian Marquardt, S. M. Girvin, and J. G. E. Harris Nature {\bf 452}, 06715 (2008).
\bibitem{jharris2} A. M. Jayich, J. C. Sankey, B. M. Zwickl, C. Yang, J. D. Thompson, S. M. Girvin, A. A. Clerk, F. Marquardt, and J. G. E. Harris, New Journal of Physics {\bf 10}, 095008 (2008).
\bibitem{Car}H. J. Carmichael, Phys. Rev. Lett. {\bf 70}, 2273 (1993).
\bibitem{Gar} C. W. Gardiner, Phys. Rev. Lett. {\bf 70}, 2269 (1993).
\bibitem{car_notes} H. Carmichael, An Open Systems Approach to Quantum Optics: Lectures Presented at the Université Libre de Bruxelles, October 28 to November 4, 1991, Springer, 1993 {\em pg. 130-132}. 
\bibitem{DSK}D.W.C. Brooks, T. Botter, T.P. Purdy, S. Schreppler, N. Brahms, and D.M. Stamper-Kurn,  Nature {\bf 488}, 476-480 (2012).
\bibitem{vanner1} M. R. Vanner et al. "Pulsed quantum optomechanics." Proceedings of the National Academy of Sciences 108.39 (2011): 16182-16187.
 
\end{references}
\end{document}